%% file: pap.tex
\documentclass[preprint2]{aastex}
\usepackage{graphicx}

\def\fs{f_{\rm S}}

\def\I0{I_{\rm 0}}
\def\tE{t_{\rm E}}
\def\t0{t_{\rm 0}}
\def\u0{u_{\rm 0}}
\def\tm4{t_{\rm 0\mu4}}
\def\te4{t_{\rm E\mu4}}
\def\um4{u_{\rm 0\mu4}}
\def\Im4{I_{\rm 0\mu4}}
\def\chim4{\chi^2_{\rm \mu4}}
\def\IS{I_{\rm S}}

\newcommand{\eg}{{e.g.},\,}
\newcommand{\ie}{{i.e.},\,}

\shorttitle{OGLE-III Microlensing Events and the Structure of the Galactic Bulge}
\shortauthors{Wyrzykowski et al.}

\begin{document}

\title{OGLE-III Microlensing Events and the Structure of the Galactic Bulge\altaffilmark{1}}
\author{{\L}ukasz Wyrzykowski\altaffilmark{2}}
\affil{Warsaw University Astronomical Observatory, Al.~Ujazdowskie~4, 00-478~Warszawa, Poland\\
  Institute of Astronomy, University of Cambridge,  Madingley~Road, Cambridge~CB3~0HA,~UK}
\email{lw@astrouw.edu.pl}
\altaffiltext{1}{Based on
    observations obtained with the 1.3~m Warsaw telescope at the Las Campanas Observatory of the Carnegie Institution for Science.}
\altaffiltext{2}{Name pronunciation: {\it Woocash Vizhikovsky}}

\author{Alicja E. Rynkiewicz, Jan Skowron, Szymon Koz{\l}owski, Andrzej Udalski, \\
Micha{\l} K. Szyma{\'n}ski, Marcin Kubiak, Igor Soszy{\'n}ski}
\affil{Warsaw University Astronomical Observatory, Al.~Ujazdowskie~4, 00-478~Warszawa, Poland}

\author{Grzegorz Pietrzy{\'n}ski}
\affil{Warsaw University Astronomical Observatory, Al.~Ujazdowskie~4, 00-478~Warszawa, Poland\\
Universidad de Concepci{\'o}n, Departamento de Astronomia, Casilla 160-C, Concepci{\'o}n, Chile}

\author{Rados{\l}aw Poleski}
\affil{Warsaw University Astronomical Observatory, Al.~Ujazdowskie~4, 00-478~Warszawa, Poland\\
Department of Astronomy, Ohio State University, 140 W. 18th Ave., Columbus, OH 43210, USA}

\author{Pawe{\l} Pietrukowicz, Micha{\l} Pawlak}
\affil{Warsaw University Astronomical Observatory, Al.~Ujazdowskie~4, 00-478~Warszawa, Poland}

\begin{abstract}
We present and study the largest and the most comprehensive catalog of microlensing events ever constructed. 
The sample of standard microlensing events comprises 3718 unique events from years 2001--2009, with 1409 not detected before in real-time by the Early Warning System of the Optical Gravitational Lensing Experiment (OGLE).
The search pipeline makes use of Machine Learning algorithms in order to help find rare phenomena among 150 million objects and derive the detection efficiency. 
Applications of the catalog can be numerous, from analyzing individual events to large statistical studies for the Galactic mass and kinematics distributions and planetary abundances. 

We derive the maps of the mean Einstein ring crossing time of events spanning 31 sq. deg. toward of the Galactic Center and compare the observed distributions with the most recent models.
We find good agreement within the observed region and we see the signature of the tilt of the bar in the microlensing data.
However, the asymmetry of the mean time-scales seems to rise more steeply than predictions, indicating either a somewhat different orientation of the bar or a larger bar width. 
The map for the events with sources in the Galactic bulge shows a dependence of the mean time-scale on the Galactic latitude, signaling  an increasing contribution from disk lenses closer to the plane, related with the height of the disk.
Our data present a perfect set for comparing and enhancing new models of the central parts of the Milky Way and creating the 3D picture of the Galaxy.
\end{abstract}

\section{Introduction}
Galactic gravitational microlensing is an astrophysical phenomenon, which originates from the fact that in the curved space-time around each sufficiently massive body (\eg stars) light travels along bent (hence converging) paths.  
The foundations of the theory lie in the General Theory of Relativity of Einstein \citep{Einstein1936} and the practical use within the scale of our own Galaxy were developed by Paczy{\'n}ski \citep{Paczynski1991} and Griest \citep{Griest1991}.
This unique astrophysical tool has numerous and very interesting applications, because, unlike other astrophysical phenomena, microlensing is sensitive to the mass of the object passing in front of a background star. 
This means that, in principle, the lens can be completely dark and it will still cause a microlensing event as long as it aligns with a background star and the relative motion between the source, lens and the observer is large enough for the event to last within the human time-scale. 
During such an event the background star appears brighter, typically by a factor of few, occasionally reaching hundreds and very rarely thousands, in the case of a perfect alignment when the images form an Einstein ring.

The fact that the microlensing is sensitive to the mass of lensing objects turns this phenomenon into a powerful probe of the mass distribution in the Galaxy, allowing for studying the structure of the Milky Way. 
\cite{KiragaPaczynski1994} defined the microlensing optical depth as a measure of the lensing probability toward the Galactic Centre and showed it as a convenient indicator of the total mass of the lensing populations along the line of sight. 
The first measurements of the optical depth obtained from the OGLE data \citep{Udalski1994} and MACHO \citep{Alcock1997} were significantly higher than expected from theoretical models of the Galaxy (\eg \cite{Dwek1995}, \cite{HanGould1995}, \cite{Binney2000}, \cite{Freudenreich1998}, \cite{EvansBelokurov2002}). 
The discrepancy was partially explained by \cite{Paczynski1994} who introduced a bar to the models, an elongated solidly revolving structure placed at about a 20 deg angle toward the Sun. 
In subsequent years the measurements of the optical depth were limited to events in which the sources were the Red Clump giants (RC) from the Galactic Bulge (\eg \citealt{Afonso2003}, \citealt{Popowski2005}).
This trick significantly moved the observed values closer to the theoretical predictions, mainly because it mostly probed a single population of sources from the bulge and reduced the problem of blending (crowding) of stars. 
Blending is a demanding issue to deal with in the extremely crowded sky regions like the Galactic Centre. 
As shown in \cite{Smith2007}, blending can not be completely neglected in the optical depth determinations, even in case of the lensing of the brightest Red Clump stars. 

Nevertheless, despite the fact that there were many hundreds of events being detected in real-time every year by the OGLE \citep{Udalski2003} and MOA \citep{Yock1998} collaborations, the number of microlensing events used for the studies of the inner parts of the Galaxy were typically much smaller. 
For instance, the optical depth measured by the EROS group relied on 16 \citep{Afonso2003} and 120 events \citep{Hamadache2006}, MACHO used 99 events \citep{Alcock2000} and 42 RC events \citep{Popowski2005}.
There were only 9 events used for computing the first ever value of the optical depth from the OGLE-I data \citep{Udalski1994}, then from the OGLE-II project 32 events were used\citep{Sumi2006}. 
The largest samples of standard single lens events used so far for the optical depth determination were 610 events from the OGLE-III years 2001-2004 \citep{WyrzykowskiPHD} and 474 events found in MOA-II in years 2006-2007 \citep{Sumi2013}.
Moreover, \cite{CalchiNovati2008} constrained the Initial Mass Function (IMF) based on just 42 events from MACHO and obtained the slope for Main Sequence stars of $\alpha_\mathrm{MS}=1.7$ and $\alpha_\mathrm{BD}=1.6$ for Brown Dwarfs populations in the Bulge, in agreement with most theoretical predictions.
However, a detailed mass spectrum can be obtained when using significantly larger samples of microlensing events.
Increasing the number of good quality and robust standard microlensing events is essential for more detailed comparison of the observations to the predictions, also over a range of Galactic coordinates \citep{Mao2012}.

The Einstein ring crossing time, \ie the time-scale of an event, is the only parameter from the standard microlensing model which has a physical meaning. 
Still, its value is a composition of multiple parameters:
\begin{equation}
\tE = \frac{\sqrt{\kappa M \pi_{\mathrm{rel}}}}{\mu_\mathrm{rel}},
\label{eq:te}
\end{equation}
where $\kappa=\frac{4G}{c^2\mathrm{AU}}\approx8.144 \frac{\mathrm{mas}}{M_\odot}$, $M$ is the mass of the lens, $\pi_{\mathrm{rel}}=1/D_d-1/D_s$, $D_d$ and $D_s$ are the distances to the deflector (lens) and the source, respectively, and $\mu_\mathrm{rel}$ is the relative proper motion between the source and the lens.
Because of that degeneracy, the time-scales can only be studied statistically in large numbers, as all parameters involved in the $\tE$ creation follow some distributions, which can be modeled for different populations within the Milky Way.
As seen in Eq. \ref{eq:te}, the strongest influence on the $\tE$ value is by $\mu_\mathrm{rel}$, therefore the distributions and dispersions of  proper motions of different Galactic populations will play the crucial role in interpreting the observed time-scales.
Measured distributions of the time-scales of microlensing events have been used in the past as one of the ways of verifying different scenarios for the composition and kinematics of the inner parts of the Galaxy, \eg 
\citet{EvansBelokurov2002}, 
\citet{BissantzGerhard2002}, 
\citet{WoodMao2005}, 
\citet{CalchiNovati2008}, 
\citet{Kerins2009}, 
\citet{Sumi2013}.

In this paper we analyze the final and complete photometric data set gathered during the OGLE-III project in years 2001-2009.
We search for high quality standard microlensing events using the optimized search criteria, supported by the Machine Learning method (Random Forest classifier), and use them to investigate the structure of the inner parts of the Milky Way.

The paper is organized as follows. 
First we describe the OGLE-III data. 
In Sec. 3 we describe in detail the procedure of the search for events, divided into several sub-steps. 
Then, we report on the results of the search and present the catalog, compare it to the EWS and discuss the detection efficiency of events.
This is followed by the discussion of the results, in which we describe the properties of the events as a whole, derive the distributions of the time-scales of events over the sky and compare it to the Galaxy models.
We summarize and conclude the paper in Sec. 6. 

\section{Data}
The data used in this work was the photometry of 150 million objects toward over 31 sq. deg. of the Galactic Bulge observed on almost 74,000 frames, \ie about 11,000 billion data points. 
We selected 91 fields out of all 177 ever observed by the Optical Gravitational Lensing Experiment (OGLE) \cite{Udalski2008} in its third phase from July 2001 until May 2009, which had a least 250 observations.
During the OGLE-III phase the Warsaw Telescope, located at the Las Campanas Observatory, Chile, operated by the Carnegie Institution of Washington (now Carnegie Institution for Science), was equipped with a mosaic CCD camera with eight 2k$\times$4k pixels detectors covering in total 0.34 square degrees.
The typical exposure time in the fields toward the Galactic Bulge was 120s allowing reaching down to nearly 21 mag in the Johnson-Cousin $I$-band, the filter in which vast majority of the observations were carried out.

\begin{figure*}
\center
\includegraphics[width=14.5cm]{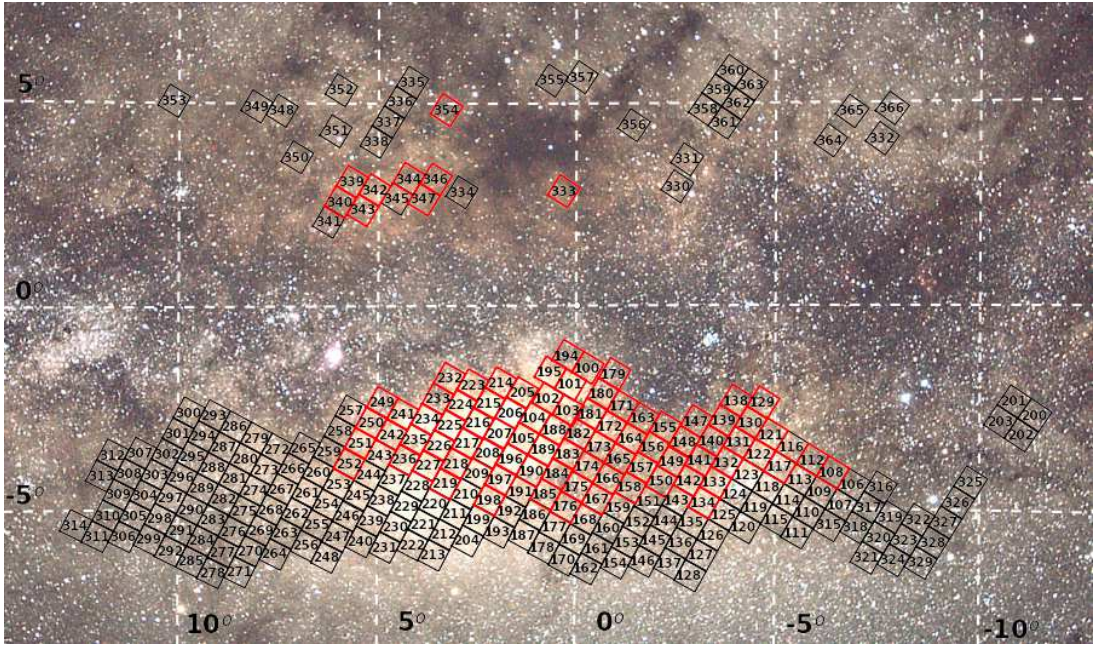}
\caption{Map of the OGLE-III bulge fields in galactic coordinates. Red squares mark fields which were used in the search for microlensing events as they had at least 250 observations over 2001-2009 period. Remaining fields were observed less frequently. Each square contains the number of the field.
The background image was taken by Krzysztof Ulaczyk.} 
\label{fig:fields}
\end{figure*}

Figure \ref{fig:fields} shows the positions of  OGLE-III fields toward the Galactic Bulge. 
Each field was observed on average once per three nights, however, from 2005 season the observing strategy changed slightly so that the most dense fields at about b$\sim$-2 deg were observed with higher cadence than the rest, typically two/three times per night.
The number of observations collected in the $I$-band over 8 years per field varied from 251 for BLG344 to 2540 for one of the central fields, BLG102.

There were typically between 1 and 35 $V$-band data points available for the fields investigated here. 
Those data were only used to obtain the averaged color of the baseline of the objects.
The calibrated color was taken from the OGLE-III Bulge photometric maps \citep{Szymanski2011}.

During the operation of the OGLE-III the data collected each night were reduced on-the-fly with the state-of-the-art difference imaging technique (DIA, \citealt{Wozniak2000}) and preliminary photometry was produced within couple of hours. 
This was the basis for the Early Warning System (EWS, \citealt{Udalski2003}), which was designed to look for new microlensing events in the real-time. 
As a result of seven years of functioning (2002-2009), the OGLE-III EWS reported about 4000 candidates for microlensing events. 

\begin{figure}
\center
\includegraphics[width=8cm]{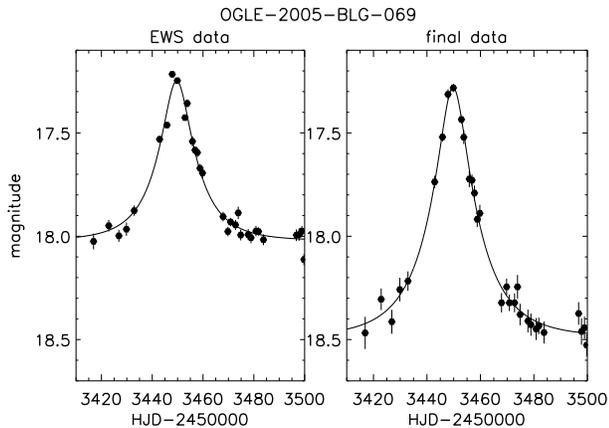}
\caption{
Comparison between the photometry data used during the real-time event search (Early Warning System,EWS) and in this work (re-reduced with new reference images) for a microlensing event found by the EWS. 
The solid lines show the best fitted microlensing models.
The scatter in the light curve decreased around the peak and the scaling factor have changed significantly in the new data. 
The measured time-scale changed from 16.3 to 19.1 days, for old and new, respectively, and the blending parameter changed from 0.48 to 0.71, suggesting that more stars were resolved on the new reference image.}
\label{fig:compareEWS}
\end{figure}

Toward the end of the OGLE-III phase, the entire observational material was re-reduced again with DIA using a new and better set of images for the composite reference images \citep{Udalski2008}, yielding significantly better quality output photometry. 
Figure \ref{fig:compareEWS} compares the quality of the photometry of a microlensing event OGLE-2005-BLG-069 reported by the EWS, obtained in the original and new reductions. 
The quality improvement is primarily due to somewhat better resolution of the new reference images.

Throughout the paper we used the re-reduced data obtained after the end of OGLE-III. 
We searched for microlensing events in this final and complete dataset of the observations of the Galactic bulge.
For the final sample of microlensing events we additionally produced a new photometry which took into account the exact position of each event on the DIA image (better than a fraction of a pixel), which in many cases yielded better quality of their light curves.

\section{Search procedure}
Extraction of rare light curves like microlensing events from vast databases is not an easy task. 
In the past, the data comprised a relatively small number of objects, making it feasible to visually inspect a small few thousands of light curves, after applying some basic selection criteria.
Nowadays, and in case of the OGLE-III data, it is necessary to seek help from automated methods of microlensing events selection.
Moreover, a fully automatized search pipeline allows for obtaining the detection efficiency of events depending on their parameters.
Machine Learning algorithms were already successfully used in time-domain astronomy, \eg \citet{BelokurovEvans2003}, \citet{Wyrzykowski2003}, \citet{Debosscher2007}, \citet{Richards2011}, \citet{Pawlak2013}, for automatization of the discovery and classification in large data sets. 

In our previous searches for microlensing events in the OGLE data in the LMC and SMC fields \citep[2010, 2011a, 2011b]{Wyrzykowski2009}, as well as in the preliminary search in the OGLE-III Bulge data \citep{WyrzykowskiPHD}, we relied on a number of cuts applied to a small number of computed features, primarily microlensing model parameters. 
In those searches the most powerful discriminator was typically the goodness of fit of the model, however, it worked fine only for the events with well observed light curves, with corrected photometric error-bars and with well understood noise. 
Therefore, such a cut could remove plausible standard microlensing events, the photometry of which could have been affected by some observational or instrumental problems, generally very common in such crowded areas as the Galactic bulge, \eg blending with variable stars.
As shown in \citet{Wyrzykowski2006} events with variability in the baseline can be useful in constraining at least some of the physical parameters of typically severely degenerated microlensing models.

Here, we developed and applied a new method for selecting microlensing events among all stars in the database. 
The database consisted of nearly a quarter of a billion of objects, therefore it required a more optimized and revised approach. 
In order to deal with such vast data set, we relied on Machine Learning (ML) techniques, in particular, on the Random Forest (RF) classifier  \citep{Breiman2001}, which finds its own the most successful multi-dimensional selection criteria. 
Like most of the ML methods, the RF requires all objects to be described in a homogenous way with various properties or features.
The advantage of the RF is that it conveniently selects which features are the most meaningful and carry the most information needed to correctly perform the classification. 
Therefore, there the risk that too many features will blur the classification outcome is minimized, which allows for preparing a large number of various features for each object.


\subsection{Preparatory steps}
Each individual light curve in the database was first pre-processed.
This included outlier removal: single data points outlying by more than 3$\sigma$ from the total mean of the entire light curve, with the a point before and after lying within 1$\sigma$. 
Then, the photometric error-bars were corrected, following the method described in \cite{Wyrzykowski2009} and \cite{SkowronPHD}, which returned a correction coefficients as a function of observed magnitude, derived from non-variable stars, taking into account all observational factors, \eg seeing and airmass.
Once the error-bars were corrected, we were able to filter out all non-variable light curves because our main goal was to find singular brightening episodes in each light curve. 
As a variability indicator for a light curve we used ratio of the standard deviation to the mean error, $\sigma_\mathrm{reltot} > 1.05$.
There were 105 million objects left from the original 150 million.
Please note, that not all of them were genuinely variable, the list included also many artifacts, \eg differential refraction effects or seeing-dependent variability caused by nearby bright and saturated stars. 

In the next step, we were looking for objects exhibiting a positive flux increase over some period of time with respect to the rest of the light curve.
For each light curve we applied a running window search for peaks, following the concept used in \cite{Sumi2006} and \cite{Wyrzykowski2009}.
In this method a window sized half of the time-span of the light curve was run over an entire light curve and for each data point $i$ a value of sigma was computed:
\begin{equation} \label{eq:signif}
\sigma_i = { I_{med,w} - I_i \over \sqrt{{\Delta I_i}^2 + {\sigma_{w}}^2 }},
\end{equation}
where $I_i$ is the magnitude of the point {\it i} and $\Delta I_i$ is
its error bar, while $I_{med,w}$ and $\sigma_{w}$ are the median and the
{\it rms} in the outer window, respectively.

Before computation we masked out all observations taken with airmass $>2$ (in most cases they were outlying due to differential refraction effects) and we averaged data points taken within the same night. 
The OGLE-III sampling was typically one observation per 2-3 nights, however, occasionally there were many observations taken within a single night, when the telescope switched to the follow-up mode to cover some interesting microlensing anomalies more densely.

From the computation of running windows for each variable light curve we obtained a set of parameters (features):
\begin{enumerate}
\item $\mathbf{N_\mathrm{peaks}}$, number of detected separate peaks, each defined as a series of points with $\sigma_\mathrm{i}>2.0$
\item {\bf maxsigma}, maximum value of $\sigma_i$ for the most pronounced peak (the highest $\sigma_i$)
\item $\mathbf{N_\mathrm{seq}}$, number of sequential data points in the most pronounced peak, all being above a threshold of $\sigma_\mathrm{thresh}=2.0$
\item {\bf peaksum}, sum of $\sigma_i$, within the most pronounced peak, \ie sum of $N_\mathrm{seq}$ sigmas.
\item {\bf peaksum7}, sum of $\sigma_i$, of the maximum and 3 adjacent points from each side of the maximum (7 points)
\item $\mathbf{\sigma_\mathrm{relB}}$, variability indicator for outside window (window B) for the most pronounced peak
\item $\mathbf{\sigma_\mathrm{reltot}}$, variability indicator for the entire light curve
\end{enumerate}


At this stage, we narrowed the sample only to those objects, which had at least 4 subsequent points in a detected peak and we were left with about 8.5 million objects.
Light curves of those were then fitted\footnote{fitting was performed using CERN's MINUIT package, http://wwwasdoc.web.cern.ch/wwwasdoc/minuit/minmain.html} with the standard \citet{Paczynski1996} microlensing model, (\ie
a point-source -- point-lens microlensing event) which is described as:

\begin{equation}
\label{eq:I}
I = \I0 - 2.5\log\left[ \fs A + (1-\fs)\right],
\end{equation}

\noindent
where

\begin{equation}
\label{eq:A}
A= { u^2 + 2 \over u\sqrt{u^2+4} } \qquad {\rm and} \qquad u= \sqrt{\u0^2 + {{(t-\t0)^2} \over {\tE^2}}}.
\end{equation}

The fitted parameters are: $\t0$ -- the time of the maximum of the peak,
$\tE$ -- the Einstein radius crossing time (event's time-scale), 
$\u0$ -- the event's impact parameter, $\I0$ -- the baseline magnitude in the $I$-band and
$\fs$ -- the blending fraction (ratio of lensed source flux to total blends' flux in the $I$-band).
The fits were performed in two ways, namely with blending parameter fixed
$\fs=1$, \ie with no blending, and with $\fs$ being free. 
For clarity, parameters of the non-blended (4 parameters) model are
given the subscript $\mu4$. 
Because the standard microlensing model is symmetrical for parameter $\u0$, we only used its positive value in further analysis.

Based on the visual inspection of a selection of variable objects from test fields BLG100.1 and BLG206.1 we derived ``common-sense cuts'' to the set of variable objects.
A cut on $N_\mathrm{seq}>=4$ was imposed to remove a vast majority of short duration artifacts, however, it  also removed poorly sampled short time-scale events, typically with time-scales shorter than 3 days. 
A dedicated search for very short duration events will be presented elsewhere.
We excluded events for which the minimization procedure returned no solution within the limited number of procedure calls, or returned unphysical values of $\tE$ and $\fs$, due to strong degeneracy between those two parameters. We further investigate the possible degeneracies by employing full MCMC modeling (see Sec. \ref{sec:catalog}).
In most cases, the lack of solution for Paczy{\'n}ski model indicated the brightening was not caused by microlensing, but was some sort of eruptive variable star.
For remaining events with converged models we required reasonable microlensing fit parameters, i.e. $|\u0|<=2$, $2150<\t0-2450000<5000$, $1<|{\te4}|<400$ days, $1<|{\tE}|<400$ days, $\fs<1.4$. 
Limits on the time of the maximum ensured we only dealt with microlensing-like episodes of brightening, where the peak time $\t0$ remained within the available data span.
The cut at 400 days in time-scale is induced by the fact of using the running-window method to find brightening episodes over light curves, in which the window size was equal to the half of the total span of the data. 
Events with time-scale of 400 days may in practice span for more than 1600 days, depending on the $\u0$, therefore, our detection efficiency for those will be smaller. 
Also, there is a large number of false events with alleged time-scale above 400 days, which turn out to be low amplitude slowly varying variable stars.
The blending parameter, $\fs$, limited at 1.4 allowed for some amount of so called ``negative'' blending \citep{Smith2007}. 
The border value was computed as a maximum value expected for crowded bulge data assuming background fluctuations due to unresolved stars at about 21 mag. 

This pre-filtering narrowed our sample to 194,000 objects. 

\subsection{Ghost events filtering}
\begin{figure}
\center
\includegraphics[width=7cm]{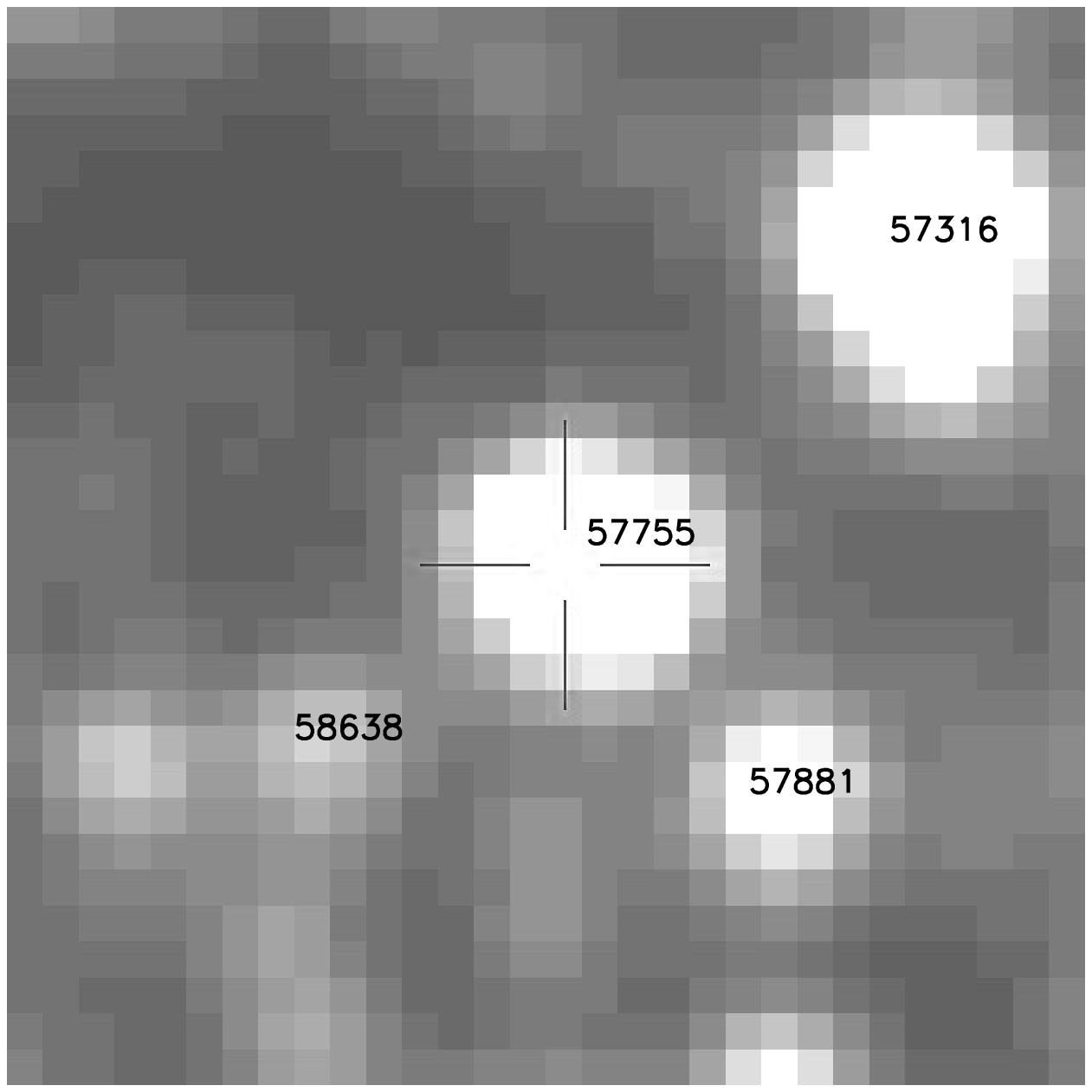}
\includegraphics[width=7.5cm]{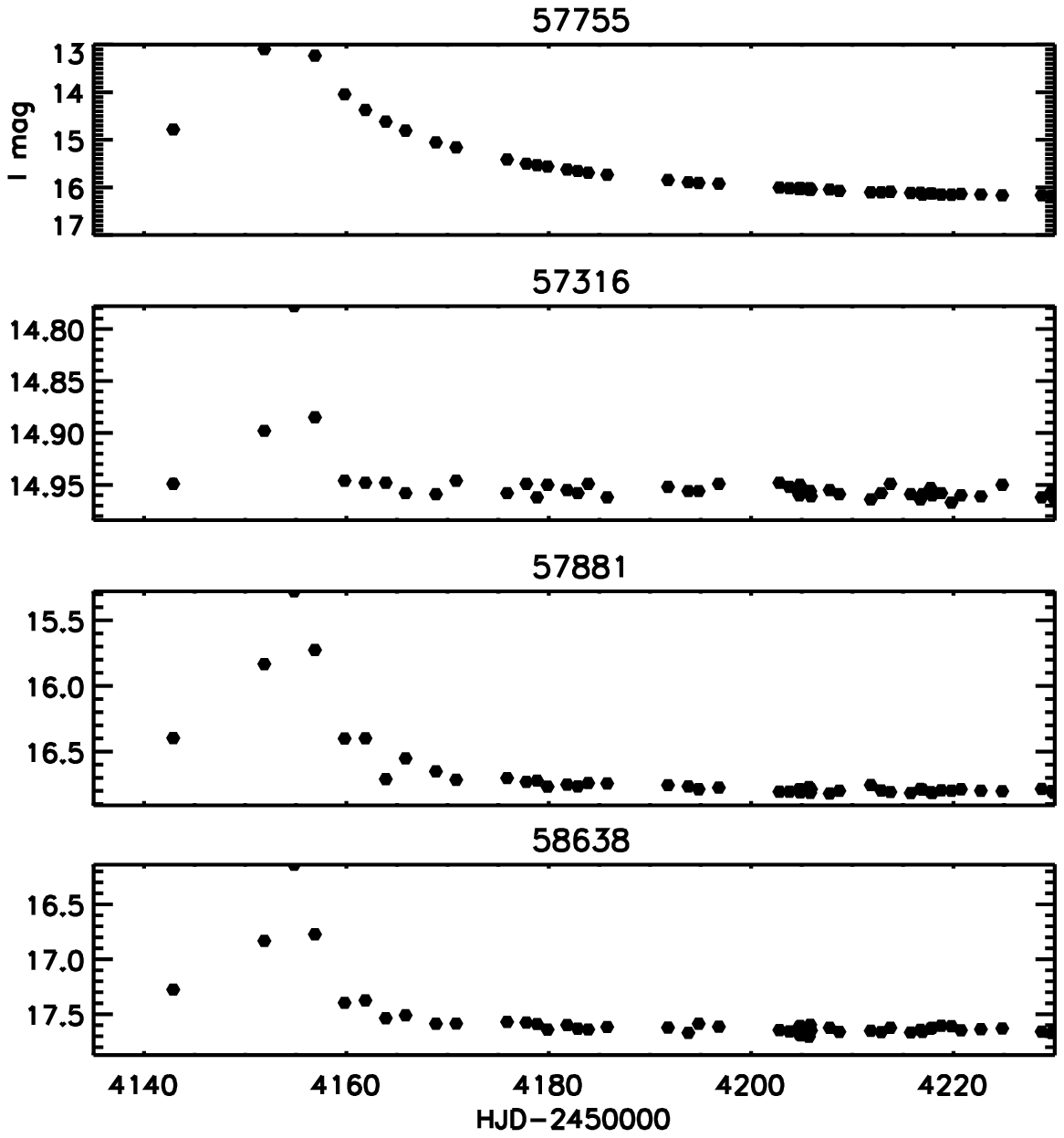}
\caption{Example of ``ghost'' microlensing events present in the database. The microlensing event occurred on star number 57755, marked with the cross on the chart sized 30$\times$30 pixels ($\sim$8$\times$8 arcseconds). 
Its light curve is shown in the top panel of the lower figure. At least three other database objects had their photometry affected by the extra flux from the brightening event. Their light curves resemble that of the ``real'' event, however the quality deteriorates with the distance from the source event. Difference imaging technique used for data reductions in the OGLE-III allows for a very accurate determination of the centroid of the microlensed flux, hence we are able to recognize which of the objects should be cataloged as a microlensing event. }
\label{fig:vst}
\end{figure}

It is very common in the crowded stellar fields that the variability of one of the stars can affect the photometry of nearby objects, often called ``ghosts'' or ``children''. 
The radius of influence depends on the seeing conditions as well as on the amplitude of the variability, therefore some of strongly amplified microlensing events can cause an effect that many nearby stars undergo a very similar brightening episode.
Figure \ref{fig:vst} shows an example of such effect in the OGLE-III data in case of a microlensing event with an amplitude of many magnitudes. 

Production of ``ghost'' events can significantly disturb the statistics and the quality of detected microlensing events. 
Therefore, it is essential to recognize which star is actually linked to the genuine microlensing event and filter out all the neighboring ghosts.
In the case shown in Fig. \ref{fig:vst} it is obvious which one is the main event, given much better quality of the light curve. 
However, not in every situation the conclusion is that simple, as in dense bulge areas there could be stars very close to each other with similar brightness, thus the wings of PSF of the additional flux may spread equally to all nearby objects and produce light curves of similar quality. 

Here we employed one of the very useful features of the difference imaging data reductions.  
Within the DIA pipeline, which was run on the entire OGLE-III data with the best set of reference images (as compared to the EWS data reductions) there is a stage at which the subtracted images are obtained and all variable sources can be identified on them with their surplus fluxes and exact positions measured, if the residuals were significant enough.
Then, in order to construct a temporal light curve, the variable source is matched to the nearest object from the reference (template) image and the difference flux is added to the flux on the reference image. 
Please note, that when the event is faint, then the pipeline uses the position of the nearest template object to measure the flux from, what in case of larger offsets can make the light curve slightly more noisy in the wings. 
The only way to improve the quality of the light curve is to force difference flux measurement at the exact position of the event. 
Such repeated forced (optimized) photometry was obtained in addition to the original light curves for the events we found, see Section \ref{sec:catalog}.

For each of 194,000 variable objects we found, we searched for the nearest differential object and read its centroid position for the frame taken at the maximum brightness of the object. 
Then if the offset from the template position was smaller than 2 pix (0.52 arc sec) we assumed that given reference object was the one with the brightening event. 
About 25\% of variable objects survived that selection and we were left only with 48,112 light curves for further classification.

With such procedure we also assured we only selected the microlensing events with a baseline, \ie we excluded all events which appeared as an excess flux with no reference star. 
Such events do exist, however they provide very limited information on source/lens and can not be used in our analysis.
Our simulation of events also excluded such events, as the events were only generated on top of cataloged stars.

\subsection{Features for classification}
For all the light curves remaining after the steps described above, we derived a set of features, used later for Random Forest classification.
First seven features were listed above and came from the running-window analysis of a light curve.

Next set of features came from the microlensing fits to each light curve:
\begin{enumerate}
\setcounter{enumi}{7}
\item $\mathbf{\te4}$, time scale of the event fitted with blending fixed to 1, describing an overall longevity of the event,
\item $\mathbf{\log{\um4}}$, logarithm of the impact parameter, sensitive to the actual height of the event,
\item $\mathbf{\chim4}$, goodness of fit of the 4-parameter model (no blending),
\item $\mathbf{\chim4/dof}$, reduced goodness of fit of the 4-parameter model,
\item $\mathbf{\tE}$, time scale of the event with free blending,
\item $\mathbf{\log{\u0}}$, logarithm of the impact parameter in 5-parameter model,
\item $\mathbf{\log{\fs}}$, logarithm of the blending parameter,
\item $\mathbf{\chi^2}$, goodness of fit of the 5-parameter model,
\item $\mathbf{\chi^2/dof}$, reduced goodness of fit of the 5-parameter model.
\end{enumerate}

For each light curve we also fitted a constant line to the out-of-event data, defined as data outside of $\tm4 \pm 5\te4$ region.
Such fit returned a mean magnitude, {\it rms} scatter (sigma) and its goodness of fit. 
To the set of features we added the following:
\begin{enumerate}
\setcounter{enumi}{16}
\item $\mathbf{\sigma_\mathrm{out}}$
\item $\mathbf{\chi^2_\mathrm{out}}$
\item $\mathbf{\chi^2_\mathrm{out}/dof}$
\end{enumerate}

The following features were then derived, using combinations of microlensing model parameters. 
They were reflecting any dramatic changes between the 4- and 5-parameter models, which often signals a non-microlensing origin of the event.
\begin{enumerate}
\setcounter{enumi}{19}
\item $\mathbf{\Delta t_\mathrm{0} = \log{|(\tm4 - \t0)|}}$,
\item $\mathbf{\Delta u_\mathrm{0} = \log{|(\um4 - \u0)|}}$,
\item $\mathbf{\Delta I_\mathrm{0} = \log{|(\Im4 - \I0)|}}$,
\item $\mathbf{\Delta \chi^2 = \chim4 - \chi^2}$.
\end{enumerate}

Microlensing modeling which gave the above mentioned parameters for each event was performed over the entire light curve. 
On top of that, we also determined the $\chi^2$ of the 5-parameter microlensing model solely for the peak and the baseline data.
Such features carried information on details of how good the microlensing model recovers the data and were sensitive to small anomalies, to which a global model was much less sensitive.
Here the peak was defined as $\t0 \pm 3\tE$ and contained $N_\mathrm{peak}$ data points.
The rest of the data (baseline) contained $N_\mathrm{base}$ data points.

\begin{enumerate}
\setcounter{enumi}{23}
\item $\mathbf{\chi^2_\mathrm{peak}/N_\mathrm{peak}}$,
\item $\mathbf{\chi^2_\mathrm{base}/N_\mathrm{base}}$,
\item $\mathbf{\chi^2_\mathrm{peak}/N_\mathrm{peak} / \chi^2_\mathrm{base}/N_\mathrm{base}}$, ratio of both $\chi^2$ values.
\end{enumerate}

The final feature to include was the $V-I$ color of the blended source, obtained from the OGLE-III photometric maps \citep{Szymanski2011}:
\begin{enumerate}
\setcounter{enumi}{26}
\item $\mathbf{V-I}$, color of the blended source.
\end{enumerate}

This set of features was derived for all light curves, which passed the basic filters described above. 

\subsection{Random Forest classifier}
For the classification of events described with 27 features derived above we decided to use the Random Forest  classifier from its Java implementation in Weka package (version 3.6.5)\footnote{\url{http://www.cs.waikato.ac.nz/ml/weka/}} developed at the University of Waikato in New Zealand. 
For training the classifier we manually selected candidate microlensing events from the pre-filtered sample of objects from the fields with various sampling properties: BLG100 (about 2400 observations), BLG180 (about 1400 observations) and BLG206 (about 1300 observations). 
During the visual classification we divided the events into three classes: {\it ULENS} for standard microlensing events, {\it EXOTIC} for non-standard microlensing events (\eg binary lens events, events with parallax effect) and {\it OTHER} for all the remaining light curves. 
Among the latter there were various types of outbursting variables like Dwarf Novae or Be-type stars, but also numerous artifactual events or other types of long-term variable stars (\eg \citealt{Soszynski2013}).
The visual selection of the events for the training set was performed independently by a few authors ({\L}W, AER and MP) in order to cross-validate the results. 
The training set was also supplemented by 30 manually selected standard events reported by the EWS.

\begin{figure}[t]
\center
\includegraphics[width=6.5cm]{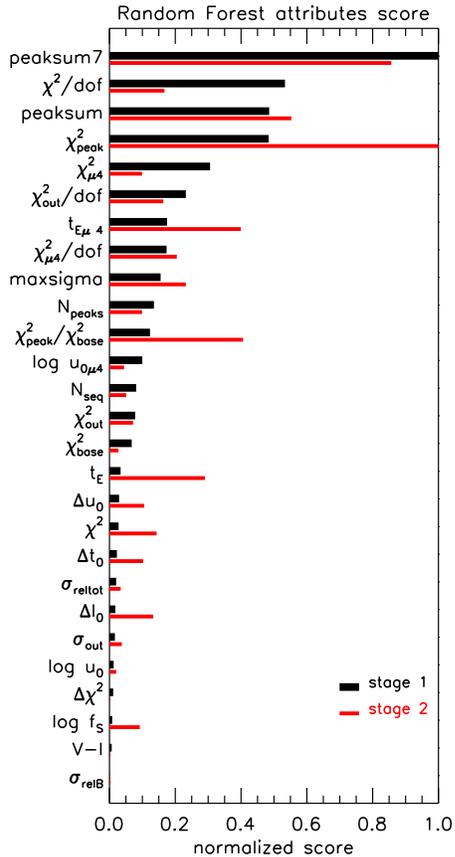}
\caption{
Score of each feature returned by the Random Forest for stage 1 and stage 2 classification. 
The scores for stage 1 are normalized to the score of {\bf peaksum7} feature, whereas scores for stage 2 are normalized to the score of {\bf $\chi^2/dof$}.  
}
\label{fig:RF}
\end{figure}

Our final training set had the following composition: {\it ULENS}: 977, {\it EXOTIC}: 53, {\it OTHER}: 2135.
The number of exotic examples was significantly smaller than the remaining classes, however, in this study we did not intend to pick the exotic events with very high effectiveness. The main reason for adding that middle class was to differentiate those type of events (typically with high signal-to-noise and large values of $\chi^2$) from standard events and other types of variables and artifacts. 
The RF classifier was setup to use 15 random features in each of the 15 decision trees, the results of which are then aggregated and the winner is chosen from the dominating class. 
The 10-fold cross-validation test of the RF classifier resulted in 96.7\% correctly classified instances (over all classes), however the false-positive rate at this stage was relatively high 27.7\%. 

Because our primary goal was to provide a sample of standard microlensing events with the lowest contamination, we added a second stage of classification to our search pipeline. 
It was designed to analyze only those events which were marked as {\it ULENS} during the first stage of the classification.
The training set comprised of only two classes: {\it GOOD} (706 instances) and {\it BAD} (271 instances), and was composed after the visual inspection of test runs of the stage 1 classifier on a fraction of the input list. 
For the verification we used field BLG104, in which we manually identified 60 genuine events, after removal of the ghost events. 
Running both stages of the RF classifier on the entire data of field BLG104 returned 56 events, with the remaining 4 being of lowest quality. 
This shows a very high efficiency of the Random Forest classifier with simultaneous high purity of the output. 
The overall false-positive rate, \ie the expected contamination for the two-stage classifier has dropped significantly to only 6.7\%.

Fig. \ref{fig:RF} shows the normalized importance score for each of the features used in the RF classifier. 
The features are sorted from the most useful during the first stage of classification at the top to the least useful at the bottom, \ie according to their importance in the decision making by the trees within the Random Forest.
The most powerful is the peaksum7 feature, which describes how strong is the event in the light curve.
Not surprisingly the second most useful feature is $\chi^2/dof$, the goodness of fit of the microlensing model.  
Also shown in this figure are the scores of the same features during the second stage. 
Because the classification here is more detailed than at stage 1, slightly different features play an important role in the classification.
For example, the most important one at stage 2 is the goodness of fit at the peak as well as its ratio to $\chi^2_\mathrm{base}$.
It shows that hierarchical division of the classification was justified as the two steps functioned on different grounds.


\begin{deluxetable}{cccclcc}
\tabletypesize{\scriptsize}
\tablecaption{OGLE-III standard events of class A.
\label{tab:events}}
\tablewidth{0pt}
\tablehead{  
\colhead{ID} & \colhead{RA$_{J2000}$} & \colhead{Dec$_{J2000}$} &   \colhead{field} & \colhead{starno} &
\colhead{EWS id} & \colhead{Duplicate}\\
\colhead{OGLEIII-ULENS-} & \colhead{[h:m:s]} & \colhead{[deg:m:s]} &   &  & &
}
\startdata
\input{table1.tex}
.... & ... & ... & ... & ... & ... & ... \\
3560 & 17:35:05.50 & -23:24:39.9 & BLG354.4 & 131 & 2002-BLG-192 & - \\
\enddata
\tablecomments{Full table in the machine-readable format is available in the on-line version of the paper.}
\end{deluxetable}

\begin{deluxetable}{rlccccccccccc}
\tabletypesize{\tiny}
\rotate
\tablecaption{MCMC microlensing parameters for 3560 standard OGLE-III microlensing events of class A.
\label{tab:mcmc}}
\tablewidth{0pt}
\tablehead{  
\colhead{ID} &   \colhead{field.starno} &
\colhead{$\t0$} & \colhead{$\tE$} & \colhead{$\u0$} &
\colhead{$\fs$} & \colhead{$\I0$} &
\colhead{$\chi^2$} & \colhead{$N_\mathrm{dof}$} &
\colhead{$V$}\\
\colhead{OGLEIII-} &  &
\colhead{HJD-2450000 [days]} & \colhead{[days]} & & & \colhead{[mag]} & & & \colhead{[mag]}\\
\colhead{ULENS-} &  & &  & & & & & &
}
\startdata
\input{table2.tex}
.... & ... & ... & ... & ... & ... & ... & ... & ... & ...\\
3560 & BLG354.4.131 & 2445.250$_{-0.074}^{+0.075}$ & 11.94$_{-0.91}^{+0.86}$ & 0.509491$_{-0.054200}^{+0.070961}$ & 1.156$_{-0.178}^{+0.263}$ & 16.693$_{-0.001}^{+0.001}$ & 401.27 & 302 & 19.137$\pm$0.024 \\
\enddata
\tablecomments{Full table in the machine-readable format is available in the on-line version of the paper.}
\end{deluxetable}

\section{Catalog}
\label{sec:catalog}

\begin{figure}[t]
\includegraphics[width=8cm]{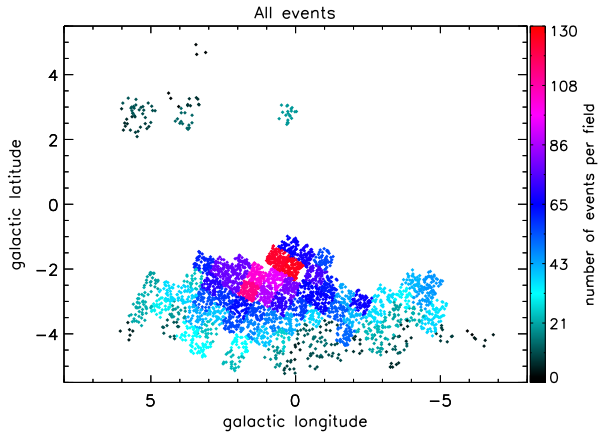}
\caption{Location of all 3560 class A standard events found in the OGLE-III Bulge data from years 2001-2009. The color of each dot indicates the number of events found in the OGLE-III field containing that event. Each field had 8 independent CCD detectors and covered 0.34 sq.deg.}
\label{fig:numberEvents}
\end{figure}

The classification procedure employing the Random Forest classifier returned exactly 3700 events. 
We then cross-matched the catalog with itself checking within 3 arc seconds radius and requiring $\t0$ to be within 10 days, and found 95 events as duplicates due to overlapping OGLE-III fields. 
We did not find any pair of events located within 3 arc seconds and which occurred at significantly different moments of time, which could have been caused by the same lensing object or a wide binary lens (\eg \citealt{Skowron+2009}).
From each pair of duplicate events we selected the event with better quality of the light curve \ie the one with higher {\bf maxsigma}, but the information on the multiplicity is stored in the table of events.
Therefore, the catalog contains 3560 unique standard microlensing events, dubbed the ``class A'' sample.
The density of events distributed over the sky in the Galactic coordinates in each OGLE-III field is shown in Figure \ref{fig:numberEvents}.
Table \ref{tab:events} lists the class A events, coordinates, OGLE-III field and star number identification, cross-match with the EWS and indication on any duplicated entries. 
The events will be labeled OGLEIII-ULENS-$nnnn$, where $nnnn$ is the number of the event, with the leading zeroes.

Each event from the catalog was fitted with the microlensing model using the Markov Chain Monte Carlo (MCMC) method\footnote{we used Python module {\it pymc} from https://pypi.python.org/pypi/pymc/}.
The priors on all model parameters were assumed uniform around the initial values, taken from the standard $\chi^2$ modeling with MINUIT code.

\begin{figure*}
\center
\includegraphics[width=8cm]{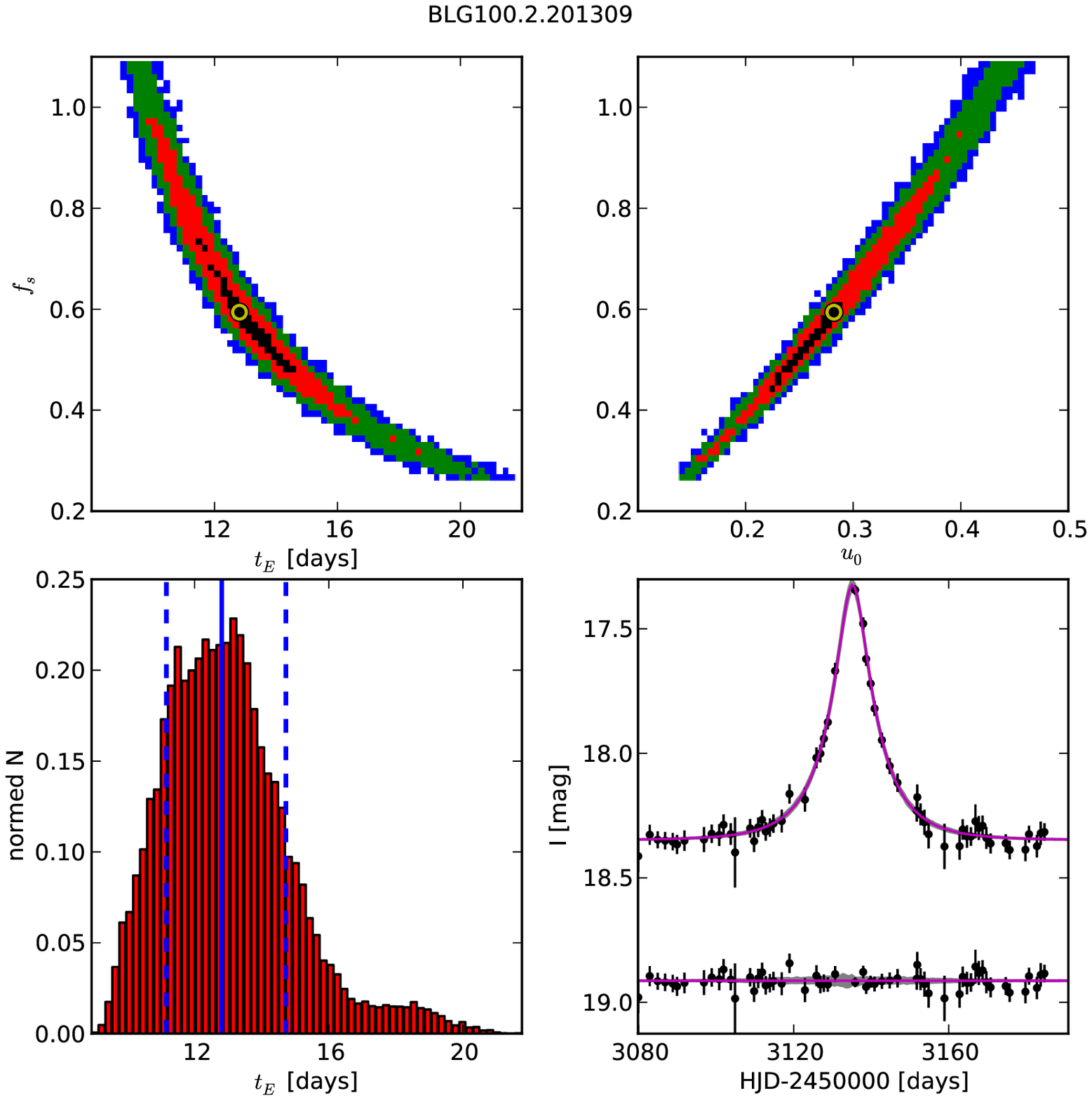}
\includegraphics[width=8cm]{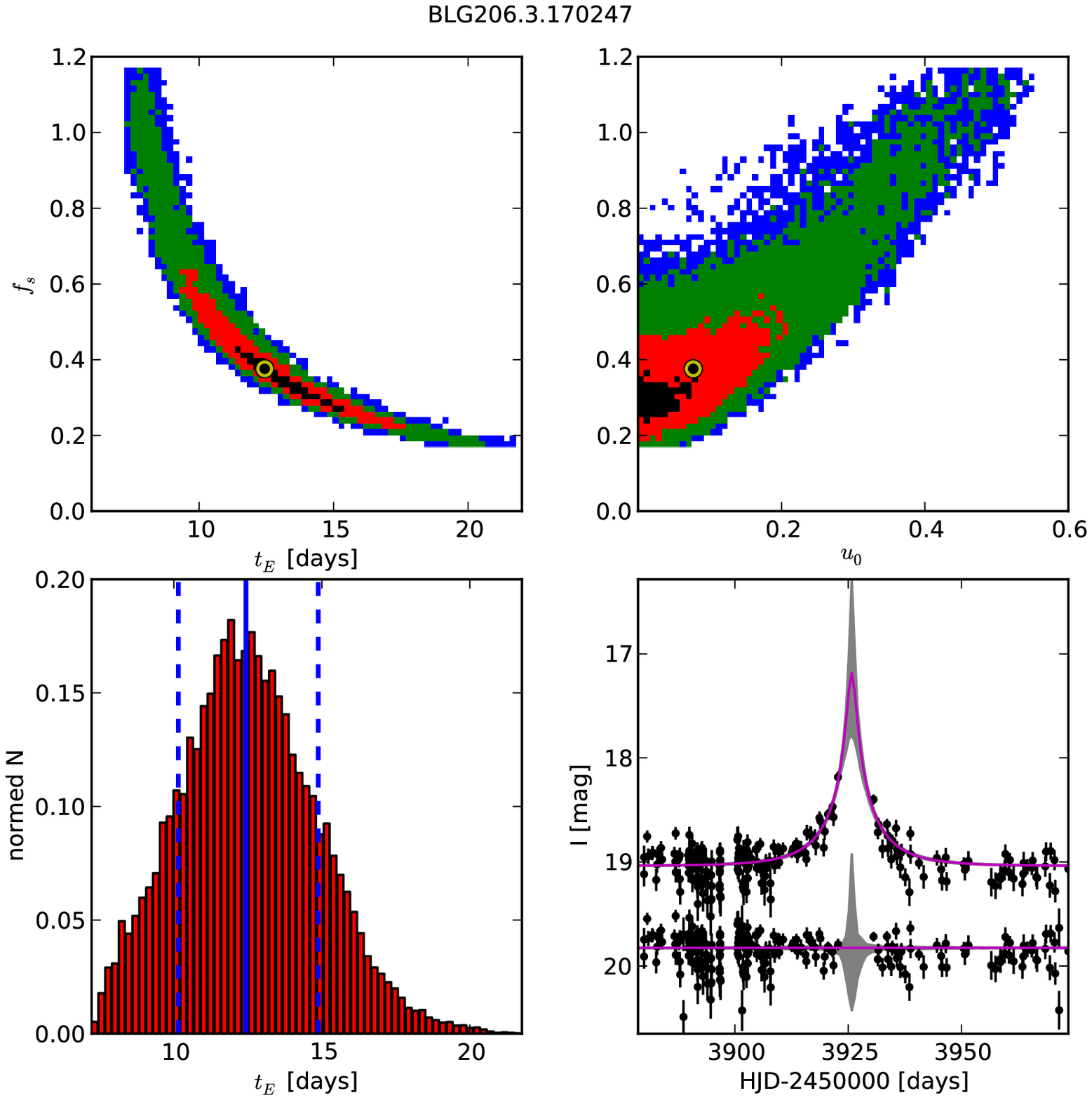}\\
\includegraphics[width=8cm]{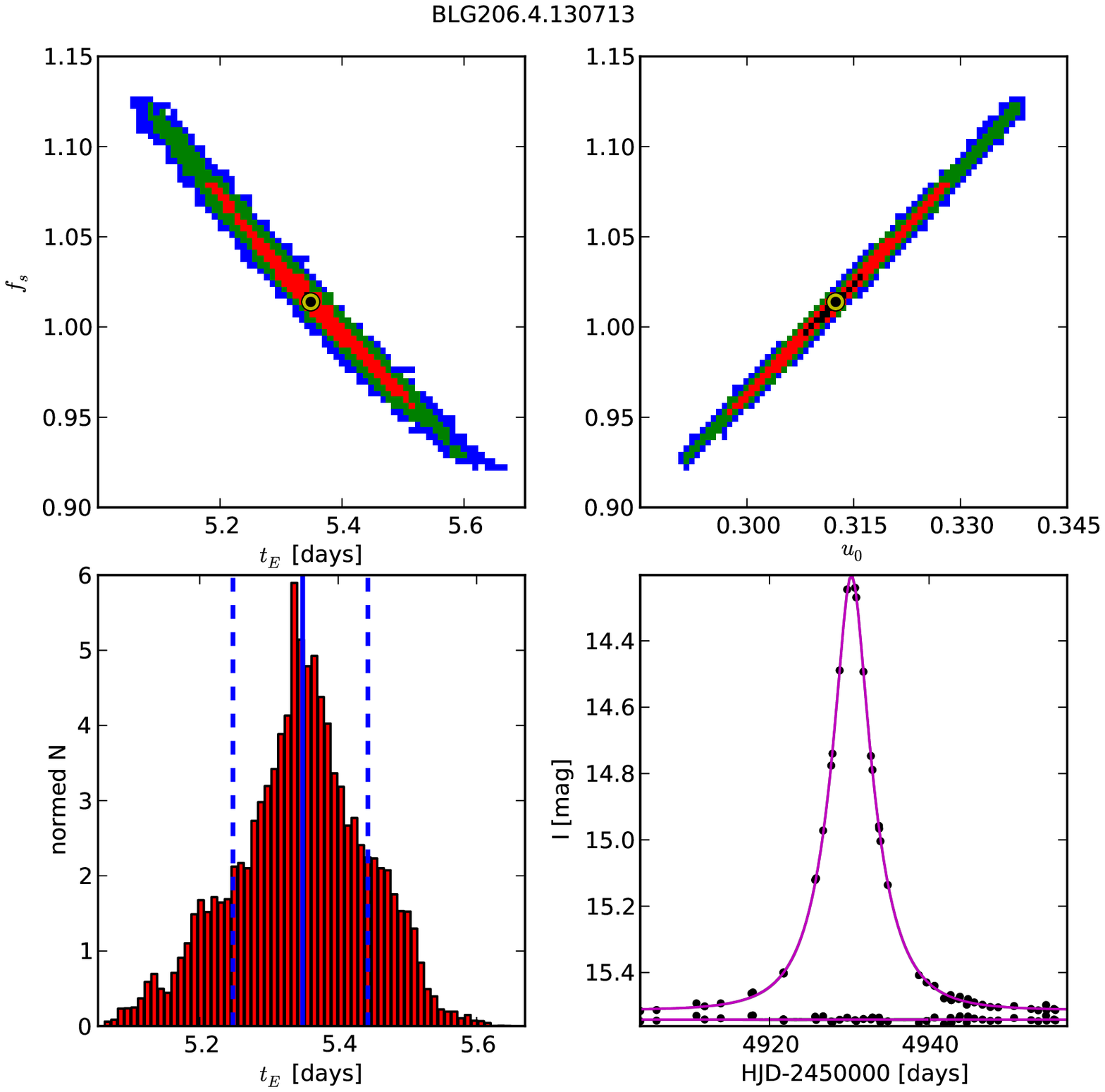}
\includegraphics[width=8cm]{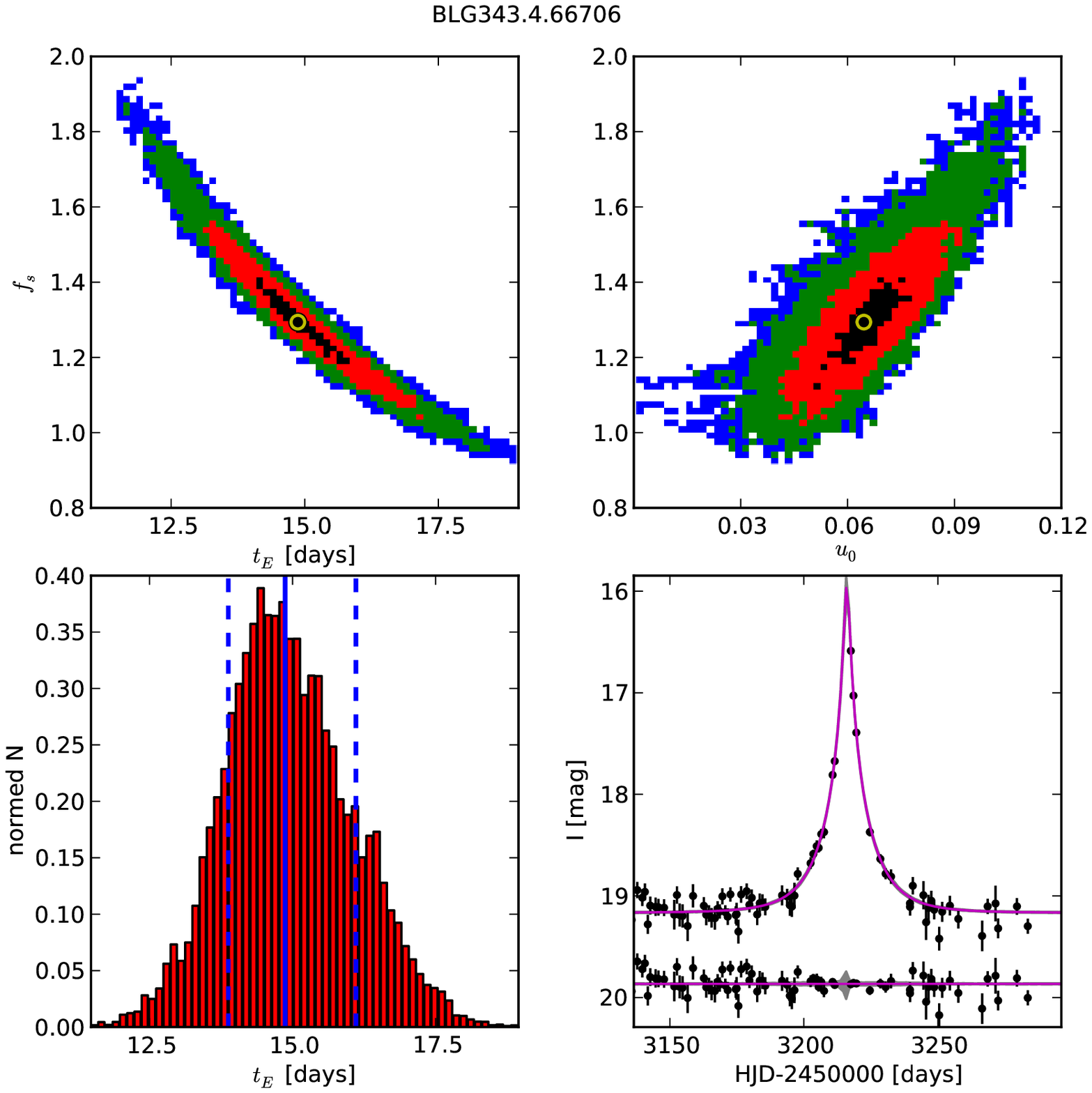}
\caption{
MCMC fitting results (density plots) for example events from the OGLE-III Bulge microlensing events catalog. 
The top panels in each figure show relations between  distributions of blending parameter ($\fs$), time-scale ($\tE$) and impact parameter ($\u0$). Colors denote sigma contours: black, red and green for 1, 2 and 3 sigmas. A magenta-yellow circle shows where the median of each distribution is located. 
Lower left panel in each of the plots is the distribution of the time-scales with the solid and two dashed lines marking the 50\%, 15\% and 85\% percentiles. Lower right panel shows the light curve of the event with the median model and its residuals (below the light curve with the offset) shown in magenta and 1-sigma models and their residuals.
Full catalog and the MCMC models for all events are available in the on-line version of the paper and on the OGLE web site.}
\label{fig:mcmcfit}
\end{figure*}

\begin{figure}[t]
\includegraphics[width=8cm]{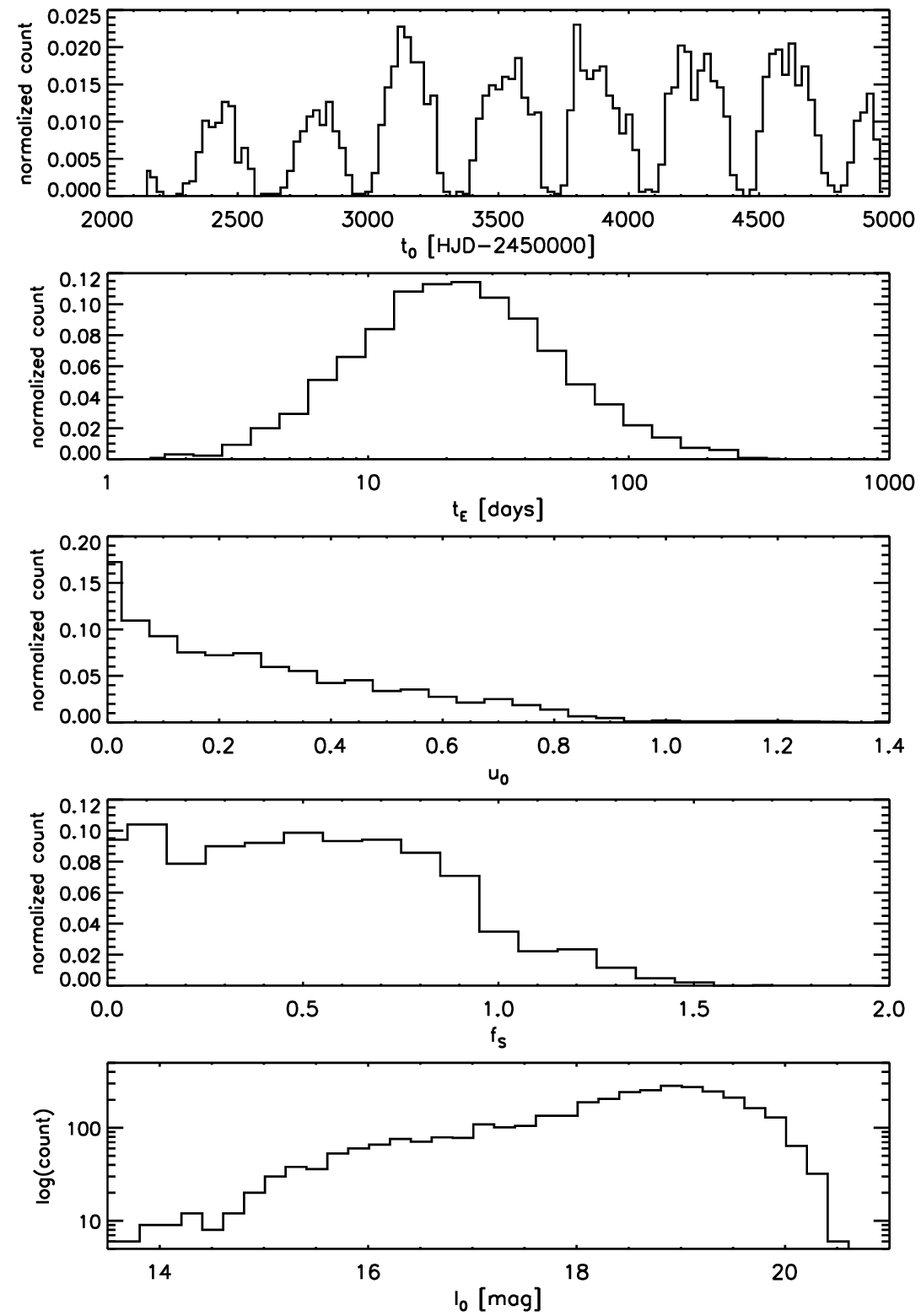}
\caption{Distributions of  microlensing parameters for the class A sample of 3560 microlensing events found in the OGLE-III bulge data.}
\label{fig:distribution}
\end{figure}

Therefore, for each event we obtained the posteriori distributions for each parameter and the median and one sigma asymmetric errors (84.1345 and 15.8655 percentiles).
Example result of the MCMC fitting to an event along with the distributions and relations for $\tE$, $\fs$ and $\u0$ are shown in Figure \ref{fig:mcmcfit}.
Such figures are available for all 3560 class A events in the on-line version of the paper, allowing for considering the degenerations of the model parameters in each of the events.
Distributions of all fitted microlensing parameters for the class A sample are shown in Fig. \ref{fig:distribution}.
Table \ref{tab:mcmc} shows all the standard Paczy{\'n}ski microlensing model parameters and their errors found in the MCMC models of the original data for events from class A.

The full machine-readable catalog, the light curves of all events and their MCMC models are available in the on-line version of the paper and from the OGLE internet archive:\\
{\it http://ogle.astrouw.edu.pl}

The light curves published there were re-processed taking the exact position of the event on the DIA image into account. That assures their best possible photometric accuracy. The photometric data were not averaged nightly, however, some cleaning was applied, similarly to the procedures described above, i.e. bad observations were masked out and the single outlying points were removed. The error-bars were also corrected.  

\subsection{Class B events}

Apart from the fully automatized search procedure to find  microlensing events, we also applied a separate search for standard events for which the full blended fit was not converging, \ie when the minimalization procedure was not finding any solutions, or was returning unphysical values for parameters, and events rejected during other stages of constructing of class A.
From a sample of light curves rejected at the pre-filtering stage we selected those for which the $\chim4/dof$ for the non-blended model was smaller than 2.5 if the baseline was fainter than 16 mag, and $\chim4/dof<5$ for $\Im4<16$.
To ensure that only high quality events passed through that filter we also required that the peak was covered with at least 10 data points and selected only events with significant amplification, \ie with $\um4<1$ in the non-blended model.

There were about 300 candidates, which were then inspected visually and an additional sample of 158 ``class B''  standard events was selected. 
Four additional events passed the criteria, but were duplicates due to overlapping OGLE-III fields.
The main reason for non-converging microlensing models for those events was in many cases some small, often gradual, long-term variability on top of the microlensing bump, likely caused by the intrinsic variability of the source or high proper motion of either the source or nearby stars.
We provide the list of those candidate events in Table \ref{tab:classB} 
and their light curves are also made available on-line for further studies.
The events from class B are dubbed following the schema: OGLEIII-ULENS-9$nnnn$, where the individual number encoded in $nnnn$ is preceded by number $9$, in order to distinguish from class A events.

Because the microlensing models of class B events were not full and did not include blending, the derived time-scales were not reliable in vast majority of the cases. 
Moreover, those events would not be detected by the automated pipeline, hence the detection efficiency did not include them and they were not used in the statistical studies below.
However, because class B events are most likely the genuine standard microlensing events, they might be a useful reference for future searches of events in their vicinity.

\begin{deluxetable}{cccclc}
\tabletypesize{\scriptsize}
\tablecaption{OGLE-III candidate standard events (class B).
\label{tab:classB}}
\tablewidth{0pt}
\tablehead{  
\colhead{ID} & \colhead{RA$_{J2000}$} & \colhead{Dec$_{J2000}$} &   \colhead{field} & \colhead{starno} &
\colhead{EWS id} \\
\colhead{OGLEIII-ULENS-} & \colhead{[h:m:s]} & \colhead{[deg:m:s]} &   &  & 
}
\startdata
\input{table3.tex}
.... & ... & ... & ... & ... & ... \\
90158 & 17:41:15.09 & -23:59:52.3 & BLG346.5 & 139563 & -\\
\enddata
\tablecomments{Full table in the machine-readable format is available in the on-line version of the paper.}
\end{deluxetable}

\subsection{Comparison with the EWS}
Comparison with the OGLE's Early Warning System (EWS\footnote{http://ogle.astrouw.edu.pl/ogle3/ews/NNNN/ews.html, where N=(2002,2003,2004,2005,2006,2007,2008,2009)}) \citep{Udalski2003} was tricky, as the reference images used for the real-time analysis and production of the final photometry we relied on had a slightly better quality and depth. 
Thus, many stars got resolved in the new data, or the blending has changed. 
Nevertheless, we compared the catalog with the events from all years of the EWS (2002-2009). 

Within 91 fields analyzed here, EWS has detected 3796 events, among which were 162 duplicates due to overlapping fields, therefore there were 3634 unique events reported.
Using a 3 arc sec matching radius and the time of maximum, we identified 2309 events in our catalog which were previously found. 
Therefore, 1409 (1333 class A + 76 class B) events are newly discovered standard microlensing events. 

The main reason for the significant majority of EWS events not recovered in our analysis was lack of convergence in the microlensing model with free blending. 
The converged modeling was among the preliminary pre-filtering requirements, \ie we requested that both blending parameter and time-scale were within sensible ranges. 
EWS was marking events exhibiting some problems with data modeling with free blending by forcing $\fs$ to be constant at 1. 
In many cases, the models did not converge because the events were not due to single-lens-single-source, or exhibited additional effects like parallax or binary lens caustics. 

The remaining EWS events which were rejected after pre-filtering at the later stage (RF classification), were inspected visually and most of them were found to be of too low signal-to-noise to be detected (\eg 2007-BLG-044).
There were also clear non-microlensing outbursts (\eg 2003-BLG-266), most likely caused by Be stars or cataclysmic variables (\eg \cite{Mroz2014}), and a significant number of exotic events with strong parallax, binary lens or finite source effects (\eg 2003-BLG-067,  2008-BLG-199).

\subsection{Detection efficiency}
The events selection procedure described above was fully automatized, therefore we were able to derive the detection efficiency for finding the standard microlensing events in the OGLE-III data.
Following the method of \citet{Wyrzykowski2009}, we simulated 300,000 light curves of events for three regions of the Bulge covered by the OGLE-III survey, with different stellar density and different sampling.
The events were simulated on top of objects cataloged on the OGLE-III reference images, such that the flux was added to the existing light curve.
We used the original database of light curves and the simulated events appeared on top of the photometry. 
Such method propagated any variability and noise properties of each light curve as well as individual measurements properties and dependence on factors like seeing, airmass and instrumental effects, which otherwise are difficult to reproduce.

In order to account for blending in the Bulge fields in our simulations, we used archival Hubble Space Telescope (HST) $I$-band images of the field BLG206 to obtain the distribution of blending parameter $\fs$. 
The distribution was derived by matching the OGLE objects to individual stars present on HST $I$-band image and by finding the relative brightness of all HST components to the OGLE brightness. 
The 91 OGLE-III fields used in this analysis were distributed over the densest region of the Bulge and, despite the changes in the stellar counts, we assumed that the overall distribution of blending parameter remained similar in all the fields. 
However, a more detailed analysis of the blending over a range of OGLE-III fields is currently ongoing as it is required to compute the real number of monitored stars, which is a necessary component of the microlensing event rate and the optical depth.

\begin{figure}[t]
\includegraphics[width=8cm]{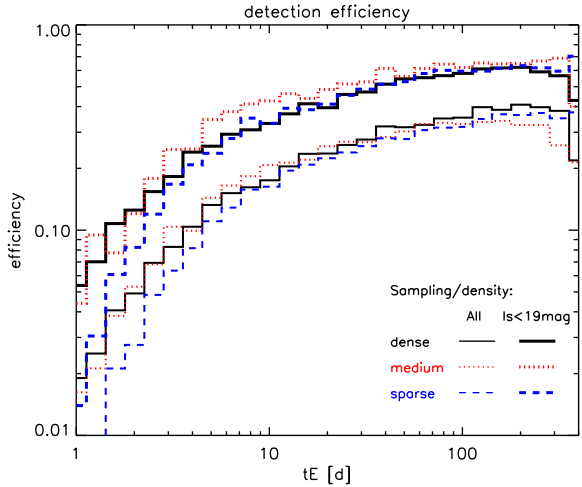}
\caption{Microlensing events detection efficiency curves as a function of simulated time-scale for all sources and for bright events ($I_S<19$mag).
Shown are the efficiencies for three fields with different sampling density.}
\label{fig:eff}
\end{figure}

For each simulated light curve we applied the pre-filtering procedure as described above, derived 27 features, and checked if the event passed through the Random Forest classifier. 
{\bf
The additional filtering of events with relative error-bar on $\tE>100\%$ was not applied in the efficiency computation as the MCMC errors were not derived for hundreds of thousands of simulated events. 
However, as noted above, this final filter only removed 2\% of events in the All Stars sample, and none of events in the bright $I_{S}<19$mag and Red Clump samples, hence the impact was negligible.
}
Figure \ref{fig:eff} shows the derived efficiency curves for three regions of the OGLE-III fields observed with different sampling: dense, medium and sparse. 
The detection efficiency for the bulk of the events is at the level of 30-40\% and drops below 10\% for events with time-scales shorter than 5 days.
At the long end the efficiency starts dropping from about 200 days to 400 days, which was the truncation in the simulations. 
The search procedure was not optimized for events with $\tE>400$ days and was susceptible to numerous low amplitude contaminants.
A dedicated search for very long events will be presented separately. 
Because the three detection efficiency curves similar, for further analysis we use the mean for all the fields.

We also derived the detection efficiency for a subsample of events, requiring the source magnitude of 19 mag criterion on simulated events. 
As expected, the efficiency is about 20\% higher than for all sources, however, the overall shape of the efficiency curve is preserved. 
The efficiencies for three different sampling (or stellar density) regions are similar to within few percent, therefore in further analysis we use the mean curve. 

\section{Discussion}

The large number of standard microlensing events allows for deriving various statistical properties of the events toward the Galactic Bulge.
For each event we derived the microlensing model parameters using the MCMC modeling and each event in class A sample is accompanied with the plot with relations between the main parameters: $\fs$, $\u0$ and $\tE$, allowing for detailed investigation of degeneracies in the models and the statistical properties of the events.

\subsection{Properties of the events}
\label{sec:properties}

\begin{figure}[t]
\includegraphics[width=8cm]{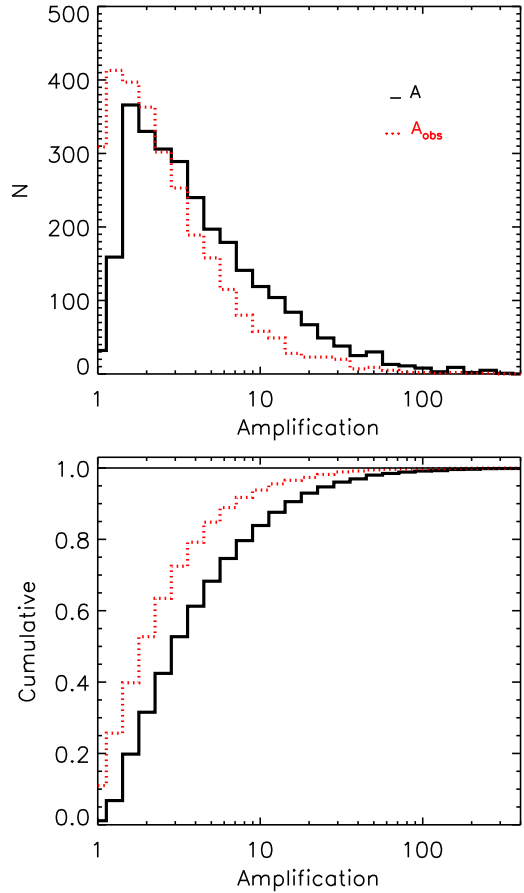}
\caption{Distributions of the maximum microlensing amplification ($A$) (solid line) and the observed amplification which takes blending into account (red dotted line). There are 24 events with $A>100$ and 5 with $A>300$. }
\label{fig:ampl}
\end{figure}

Figure \ref{fig:ampl} shows distributions of maximum amplifications for 2812 events selected from the entire sample to have well constrained impact parameter, \ie $\Delta\u0/\u0 < 1$. 
The drop between 1 and 2 is caused by the lower detection efficiency for events which were not magnified significantly above the noise. 
There are 32 events with amplification greater than 100 and 6 with amplification greater than 300. 
The cumulative histogram on Fig. \ref{fig:ampl} shows that half of the events had $A>3$. 
Also shown (red dotted line) is the distribution of the observed amplification, $A_\mathrm{obs}=(A-1)\fs+1$, \ie the actual observed rise in brightness given the blending. There were 8 and 1 events which got 100 and 300 times brighter, respectively, typically with $\fs\sim 1$.
Half of the events had their observed magnification greater than 2.0. 
The maximum amplification for simulated events presented a very similar distribution, indicating self-consistency between data and simulations.

\begin{figure}[t]
\includegraphics[width=7.0cm]{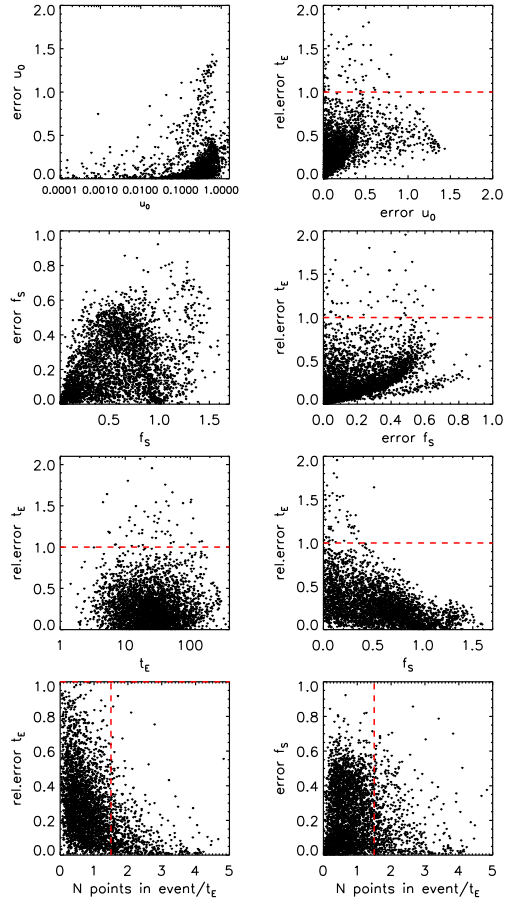}
\caption{Relative errors on the main microlensing parameters: $\u0$, $\fs$ and $\tE$. 
Despite the error in determination of $\u0$ and $\fs$ spans a very wide range, well above 100\%, the error on $\tE$ stays mostly within 100\%. 
Only 2\% of events have relative error in $\tE$ larger than 1.  
}
\label{fig:params-errors}
\end{figure}

Figure \ref{fig:params-errors} shows how the MCMC errors (approximated here as symmetric) of microlensing parameters are spread.
The events with poorly constrained amplification are forming a separate group on most of the panels, however, it can be seen that their time-scales are still well recovered, with relative error on $\tE$ below 1.
Only about 2\% of events have the relative errors in $\tE$ greater than 100\% and about 17\% have it larger than 50\%.
This allows to select a subsample of events with well constrained Einstein radius crossing time, the events which can then be used for statistical studies, computing the optical depth or constraining the shape of the IMF. 

We notice that larger relative error on $\fs$ and $\u0$ does not necessarily lead to a bad derivation of the $\tE$, however, it is clear that majority of badly constrained time-scales are related to $\fs<0.5$, \ie where the blending is severe.
Strong blending causes the light curve to show only the tip of the actual event when the observed amplification becomes high enough to overcome the noise of the baseline. 

We also find that vast majority of poorly constrained time-scales is related to the sparse sampling and is not to the time-scale itself.
The bottom panels of Fig. \ref{fig:params-errors} show the dependence of the error on $\tE$ and $\fs$ as a function of the average number of data points taken per 1 day unit of the time-scale of the event. 
It can be seen that all events with more than 1.5$\times\tE$ points collected had their time-scales derived with small relative error. 
This property can be useful when designing future microlensing experiments, however, we emphasize that is only valid for the time-scales longer than about $\tE>5$ days, as there were not enough events below that range in our sample.
Determination of the blending parameter is somewhat less sensitive to the average number of points taken during the event, however, it depends strongly on the coverage of the wings of the event, as shown already in \cite{WozniakBlending}.

\begin{figure}[t]
\includegraphics[width=7cm]{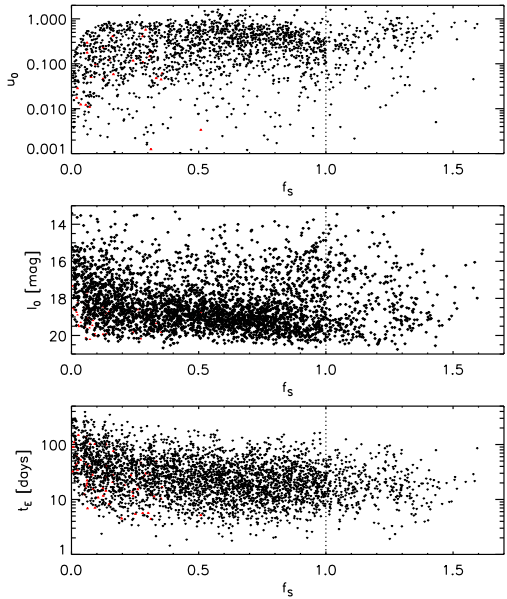}
\caption{Dependence of $\u0$, $\I0$ and $\tE$ on the blending parameter $\fs$, as obtained in the MCMC modeling of the microlensing events. 
Red dots point out events for which the relative error on $\tE$ was larger than 100\% and which were excluded from time-scale distribution analysis.
}
\label{fig:blending-relations}
\end{figure}

Figure \ref{fig:blending-relations} shows blending parameter distribution with relation to $\u0$, $\I0$ and $\tE$. 
We clearly see the observational bias that highly blended events ($\fs \ll 1$) can only be detected if highly magnified, \ie with very small $\u0$. 
Then, the observed (blended) baseline magnitude follows roughly the results of the blending simulations from \citet{Smith2007}.
Agreement with simulations indicates the blending parameter obtained for in our microlensing models follow the expected distribution, which, in turns, means we derive the time-scales of events correctly.
The relation shows that for bright events blending tends to be around 1, however can also take much smaller values.
As already pointed by \citet{Smith2007}, events with bright baseline can not be assumed to suffer less blending than fainter events.
For fainter events, below 18 mag, the blending parameter can be of any value. 
Events with relative errors on time-scale larger than 100\% tend to have smaller $\fs$, indicating the blending uncertainty is mainly responsible for larger errors in $\tE$.

There is a signature visible on the $\tE$-$\fs$ plot that a fraction of long events tends to have small $\fs$. 
This could be an artifact of modeling and can lead to a bias in derived distributions of $\tE$ toward longer time-scales. 
However, as we show later, we found a way to avoid such biases.
The same plot also shows that there is hardly any event with $\tE$ shorter than 15 days as very small $\fs$, which, in turn, is expected, as severely blended short time-scale events have small $A_\mathrm{obs}$ and hence small detection efficiency.

\subsection{Time-scale distribution}
One of the main applications for a large ensemble of microlensing events is studying the structure of the inner parts of the Galaxy.
This is usually done with the microlensing events rate ($\Gamma$) and the optical depth ($\tau$) (\eg \citealt{EvansBelokurov2002}, \citealt{Mao2012}).
Both of those quantities rely on the time-scales, however, they also require the knowledge of the number of monitored stars. 
As noted in \eg \citet{Sumi2006} and \citet{Wyrzykowski2009}, obtaining that value unambiguously  is not straightforward for the dense stellar fields.
Microlensing can occur on stars which are very faint and are blended with brighter foreground stars, therefore in order to know the exact number of monitored stars it is necessary to understand the blending in observed fields, typicaly by comparing the observed luminosity function (LF) to the one from the images of much higher resolution (\eg from the HST) in which the individual stars are resolved. 
In \citet{Wyrzykowski2011a} and \citet{Wyrzykowski2011b} such analysis 
was performed for the OGLE-III data of the LMC and the SMC, assuming no major variations in the LF between fields.
However, repeating it for the Bulge data is much more complicated. 
OGLE-III covered a wide area of the sky toward the Galactic Center, where the extinction and the mix of stellar populations vary significantly from field to field. 
A detailed analysis of the number of stars and blending distribution over a range of OGLE-III fields will be performed in near future, here however, we still can use the observed time-scales, corrected for the detection efficiency, to investigate the effects of the structure of the Galaxy on the distribution of $\tE$. 

\subsubsection{Comparison to Besan\c{c}on model}

The most recent model of the Galaxy and its microlensing yield in the Bulge \citep{Kerins2009} derived the mean time-scale  for events which occurred on sources with resolved magnitude brighter than $\IS<19$ mag. 
From sample A of our catalog we selected 1205 events for which the source magnitude (computed using $\I0$ and $\fs$) was brighter than 19 mag and the relative error on $\tE$ was less than 100\%.
We binned the events into $1\times 1$ deg$^2$ bins, 
requiring at least 5 events in a bin -- most bins contained actually about 30 events, with the maximum of about 100 in the densest area.
Table \ref{tab:tebins} contains the values of mean time-scale and the number of events in each bin.

\begin{figure*}
\center
\includegraphics[width=12cm]{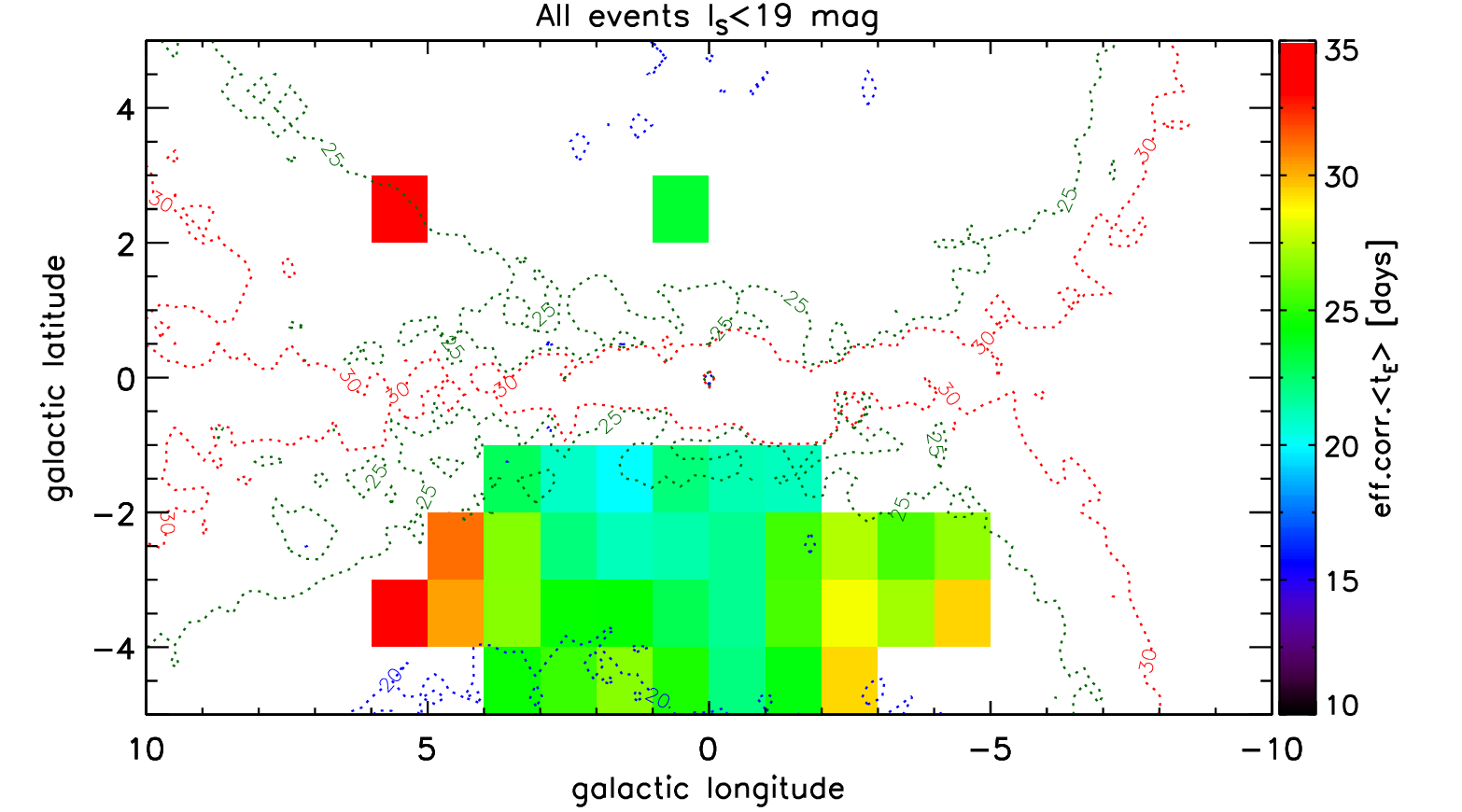}\\
\includegraphics[width=12cm]{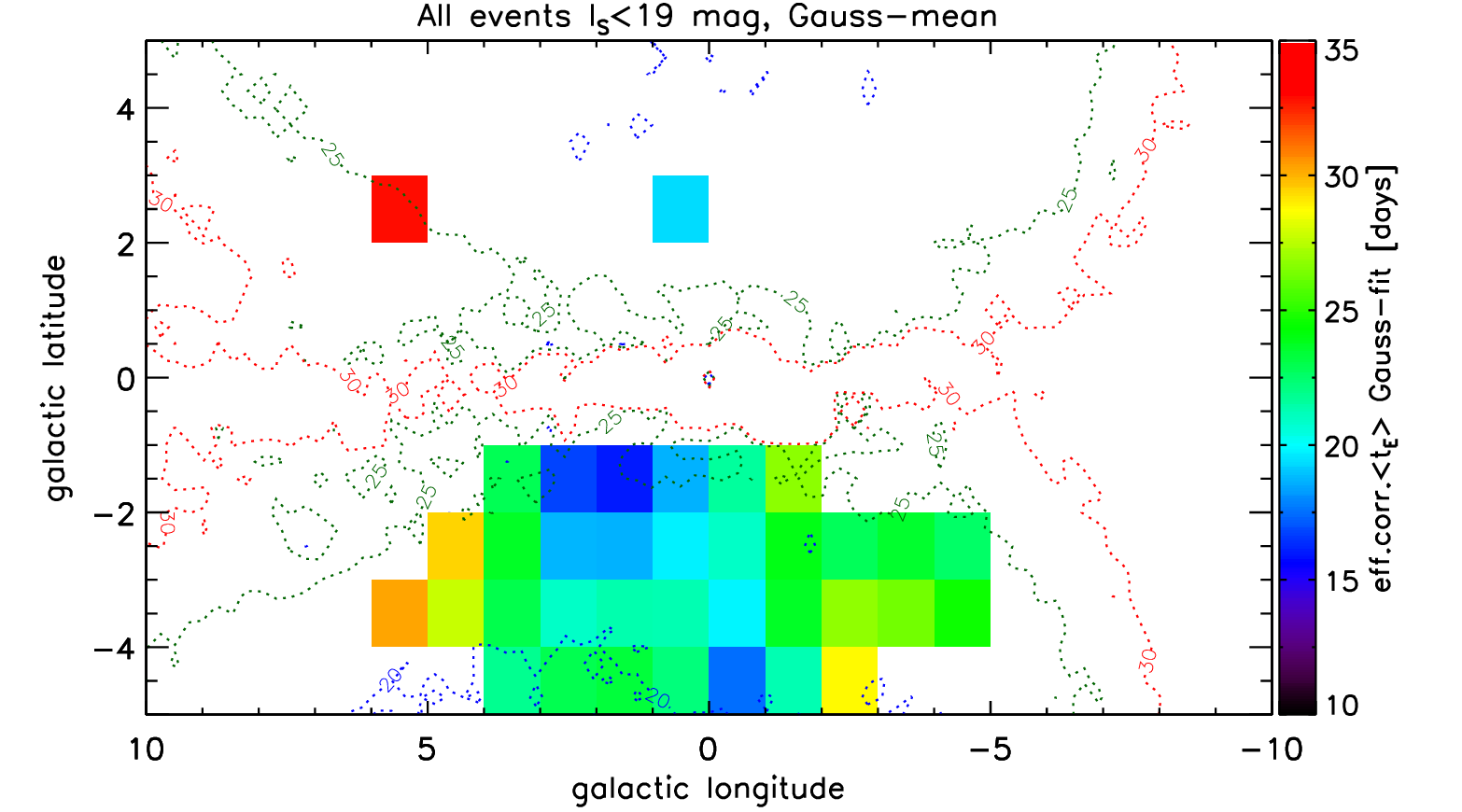}
\caption{
Efficiency corrected mean $\tE$ map for 1205 events with source(resolved) magnitude brighter than 19 mag and relative error on the time-scale better than 100\%.
Dashed contours show the expected mean time-scale (at 20, 25 and 30 days) as computed in \cite{Kerins2009}. 
Within each bin (size 1$\times$1 degree) the mean time-scale was computed from the actual values of $\tE$ (upper map) and from the log-normal fit to the distribution (lower map). Both maps were smoothed with a Gaussian with FWHM=1.75 deg.
}
\label{fig:temap19}
\end{figure*}

Within each bin we created a distribution of $\tE$ in log-space, convolved it with the detection efficiency for bright sources, and computed the arithmetic mean.
Additionally, in order to minimize the effects of single long- or short-time-scale events affecting the computation of the mean $\tE$, we also fitted the distribution in each bin with a log-normal model, and then computed the mean for that model.
Figure \ref{fig:temap19} shows maps of the mean time-scale obtained in those two ways.
The maps are also overlaid with the expectations for the Besan\c{c}on model from \citet{Kerins2009}.

Computation of the mean time-scale based on the log-normal model clearly is less prone to outliers in the $\tE$ distribution, as can be seen in the bins at high galactic longitudes, where there were very few events in each bin. 
The averages computed from a log-normal fit are somewhat smaller in most of the bins, also proving that the regular mean is affected by long events. 

\begin{deluxetable}{cccccc}
\tabletypesize{\footnotesize}
\tablecaption{Mean time-scale bins for OGLE-III events.\label{tab:tebins}}
\tablewidth{0pt}
\tablehead{\colhead{$l_\mathrm{central}$} & \colhead{$b_\mathrm{central}$} & \colhead{$\langle \tE \rangle_\mathrm{uncorr}$} &   \colhead{$\langle \tE \rangle_\mathrm{eff.corr}$} & \colhead{$\langle \tE \rangle_\mathrm{Gauss}$} & \colhead{$N_\mathrm{events}$}\\
\colhead{[deg]} & \colhead{[deg]} & \colhead{[days]} &   \colhead{[days]} & \colhead{[days]} & \colhead{}}
\startdata
\input{table4.tex}
\enddata
\tablecomments{Machine-readable table is available in the on-line version of the paper.}
\end{deluxetable}

As can be seen in Fig. \ref{fig:temap19}, OGLE-III events are located almost solely between two isolines of the \cite{Kerins2009} model, between 20 and 25 days. 
However, in the central parts, at $l\sim0$, our values tend to be significantly below 20 days, on both maps. 
The mean time-scale then increases with increasing $|l|$, but with clear asymmetry and larger values on the negative galactic longitudes.
This is most likely a signature of the bar geometry, also present as an asymmetry on the synthetic map of \citet{Kerins2009}.
-- the fraction of bar-bar and bar-disk events is different depending on the view angle at the bar \citep{Paczynski1994}.

\begin{figure*}
\center
\includegraphics[width=13cm]{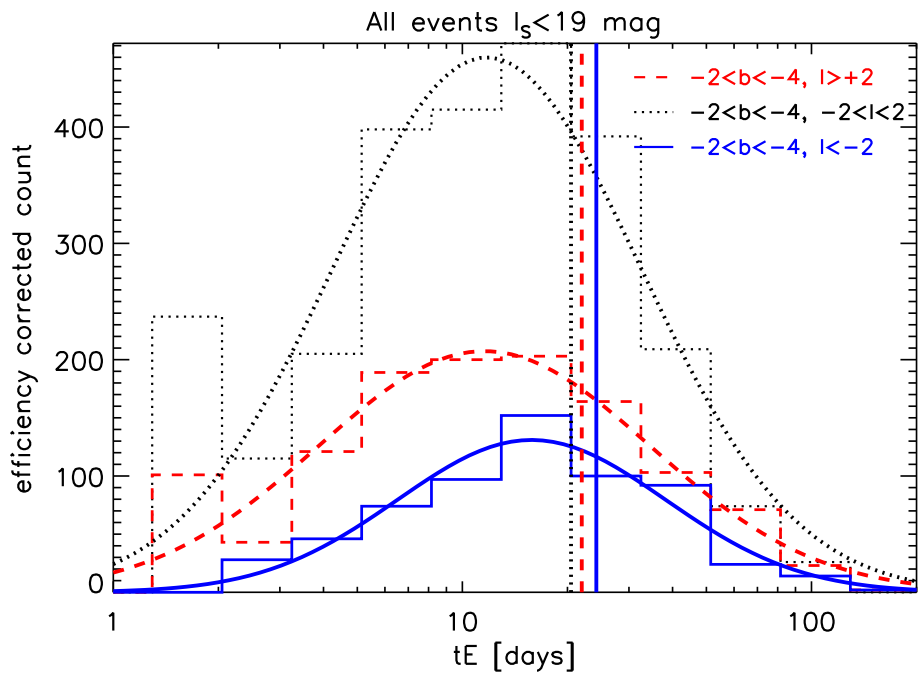}
\caption{Distribution of the time-scales of events for three regions between -2 and -4 galactic latitudes: positive galactic longitudes (dashed red), central (dotted black) and negative (solid blue). 
Only events with $I_S<19$ mag and relative error on $\tE<100\%$ are used for the histograms. 
Also shown are the best fit log-normal models with the vertical lines showing their mean values.
The highest mean is measured as 24.2 days for the negative bins (solid blue).
}
\label{fig:histlogte-bar}
\end{figure*}

The effect of asymmetry is also visible if the events are binned into three broad regions: for positive, central and negative galactic longitudes, within the same galactic latitude band (from -4 to -2 degrees). 
The distributions for those regions are shown in figure \ref{fig:histlogte-bar} along with the best fit log-normal models.
The standard mean $\tE$ (computed from the actual distribution) for positive and central regions is 23.2$\pm$0.7 days and 20.3$\pm$0.4 days, respectively. 
For the negative longitudes bin the mean time-scale is 27.9$\pm$1.2 days, which clearly stands out with respect to the other bins.
The log-normal model mean $\tE$ values for positive and central bins is 22.0 and 20.5 days, respectively.
For the negative bins it is 24.2 days, somewhat smaller than the standard mean, however, still clearly higher than the mean time-scale in positive and central parts of the map.
Moreover, the mean time-scales for negative galactic longitudes are higher than expected for the Galactic model of \cite{Kerins2009}. 

OGLE-III fields do not cover the regions beyond $|l|>5$, where the increase in the mean $\tE$ should be even more pronounced, however, the fact that the mean time-scale increases somewhat quicker with galactic longitude may indicate either that the bar is tilted somewhat more toward the line-of-sight. 
Another reason of such rise could be that the boxy bar is wider than assumed in the Besan\c{c}on model, producing more bar-bar events at lower $l$ than in the center.
A future comprehensive analysis of the OGLE-IV survey data, which is monitoring the region up to $|l|=10$ since 2010, will give even more clues to verify the shape of the galactic bar.

When comparing the mean time-scale structure to the theoretical predictions from \cite{EvansBelokurov2002} for the FreudenreichÕs bar model (their Fig. 5), it resembles the model without bar streaming included.
On the other hand, the values of the time-scales here are from 50 to 75\% larger than predicted in \cite{EvansBelokurov2002}, but still significantly lower than those expected for bar streaming. 

\subsubsection{All events sample}
\begin{figure}[htb]
\includegraphics[width=7.8cm]{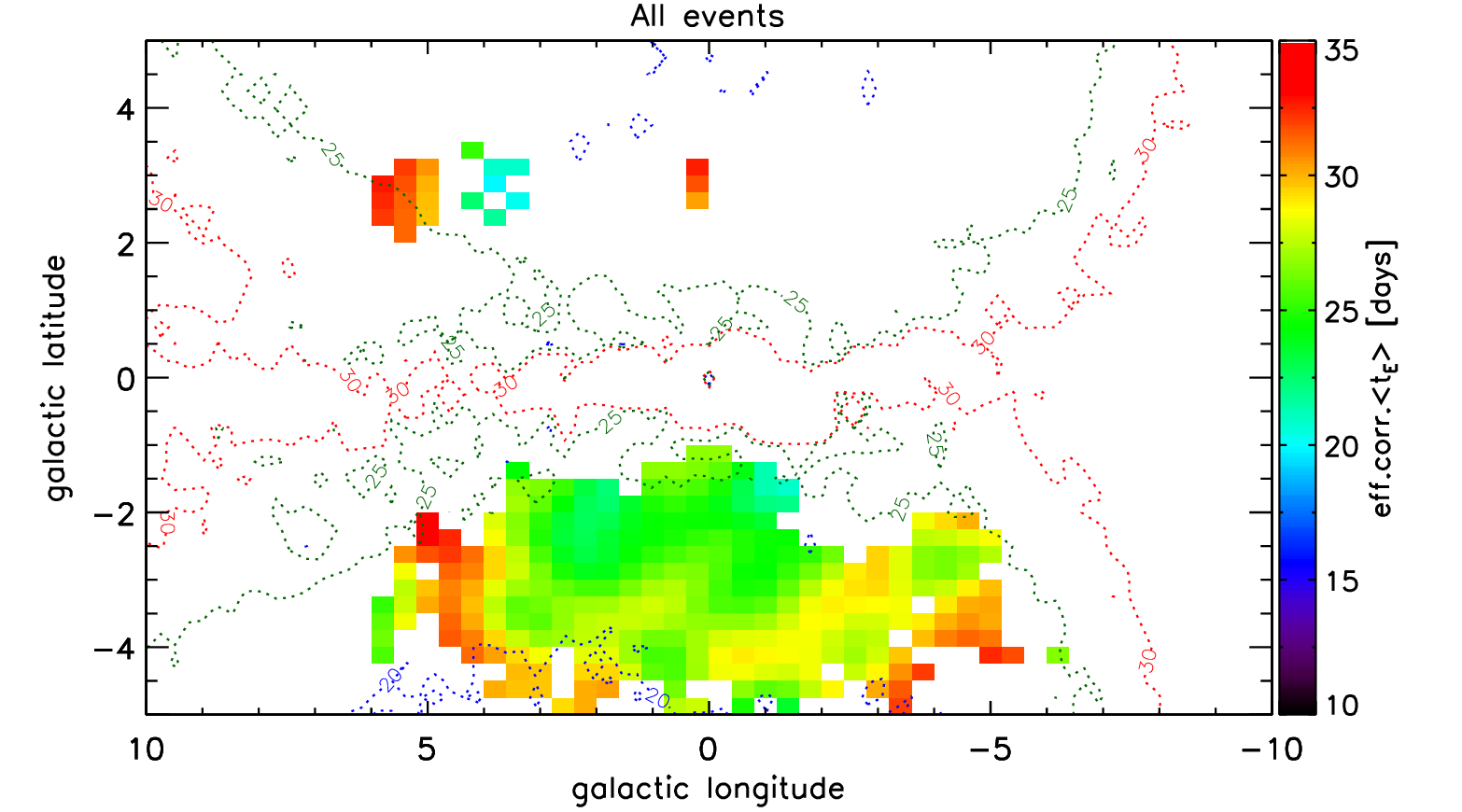}\\
\includegraphics[width=7.8cm]{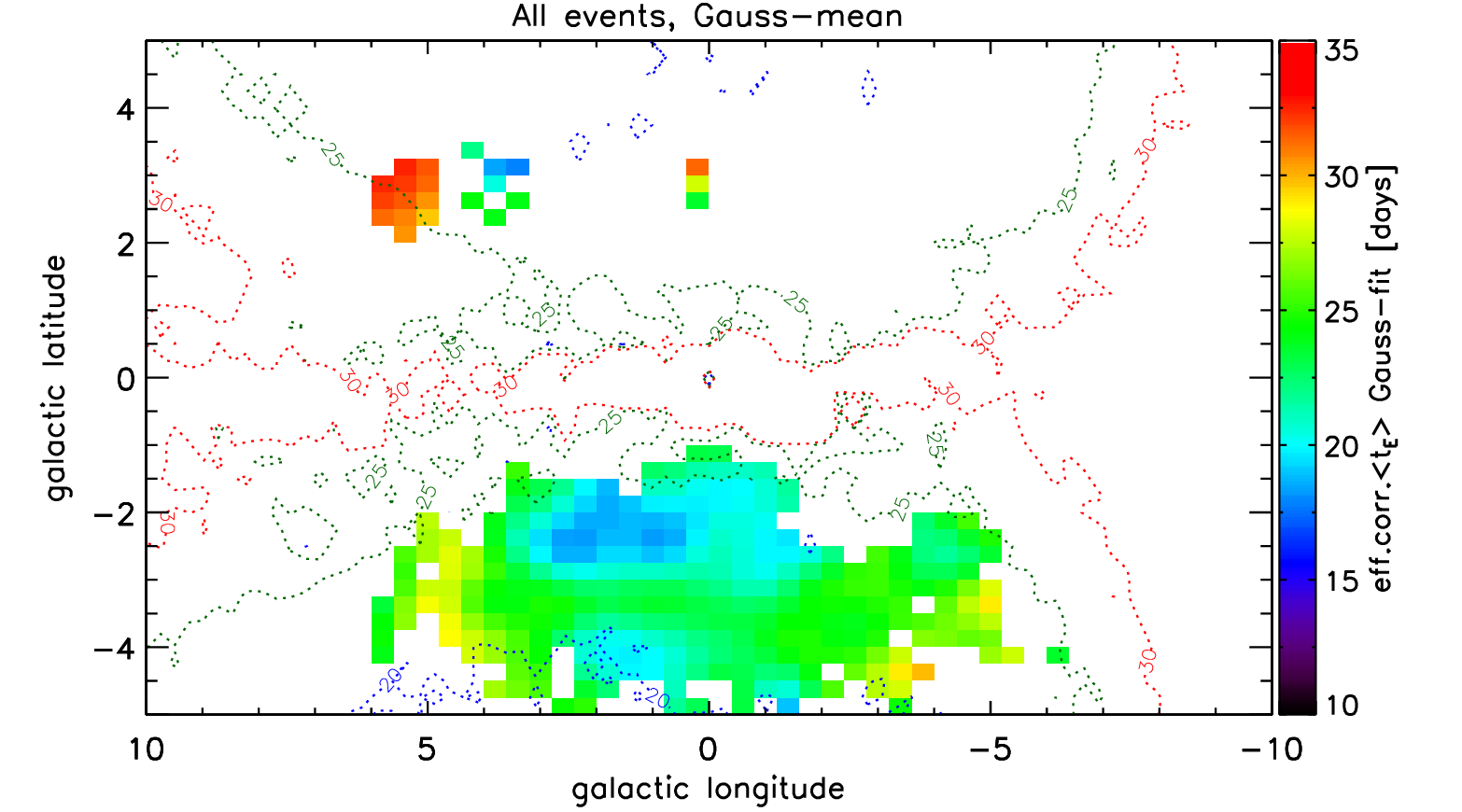}
\caption{
Maps of efficiency corrected mean $\tE$ from computed in standard way (top) and based on log-normal model fit to the $\tE$ distribution in each bin (bottom).
Maps were build for 3500 events with no restriction to the source magnitude and with the relative error of $\tE$ better than 100\%.
The bins are 0.4$\times$0.25 degrees and the maps are smoothed with a Gaussian with FWHM= 3 $\times$ bin size. 
Dashed contours show the expected mean time-scale for events with $\IS<19$ mag \citep{Kerins2009}, shown here for reference only.
}
\label{fig:temapALL}
\end{figure}

Figure \ref{fig:temapALL} shows maps of mean time-scales computed in a standard way and from the log-normal-fit for all class A events with relative error in $\tE$ better than 100\%, with no restriction regarding their source or baseline magnitude. 

There were 3019 such events and the maps were binned with small bins of 0.4$\times$0.25 degrees, a similar size as shown on map of mean time-scales obtained from MOA-II data \citep{Sumi2013}. 
The main difference between standard and log-normal-fit mean time-scales is clearly visible again on the edges of the observed region, where there were very few events (even just 2 per bin) to compute the $\langle \tE \rangle$ and the arithmetic average was very sensitive to long events. 
There is some broad structure visible again on both maps, with the lower mean time-scale values on the positive galactic longitudes and higher on the negative longitudes, however, the finer elements of the map are of very low statistical significance due to low number of events in each bin.
The all-events sample is a mixture of events from various combinations of populations: bar-bar, disk-bar and disk-disk, therefore the observed distributions are generally more blurred, as different pairs cause different duration events, with bar-disk being typically the shortest, whereas bar-bar and disk-disk the longest.

The overall log-normal-fit mean time-scale still stays within predicted 25 days in the central parts, but for large $|l|$ reaches close to 30 days.
This map could be compared to Fig.4 of \citet{EvansBelokurov2002} for Freudenreich's model \citep{Freudenreich1998}, which shows the mean time-scale for sources and lenses from either disk or the bar, however due to different resolutions it is difficult to conclude on the comparison.
Nevertheless, the values on our map tend to agree with the expectations for the model with the contribution of the spiral structure for the Freudenreich's model.

\begin{figure}
\includegraphics[width=8cm]{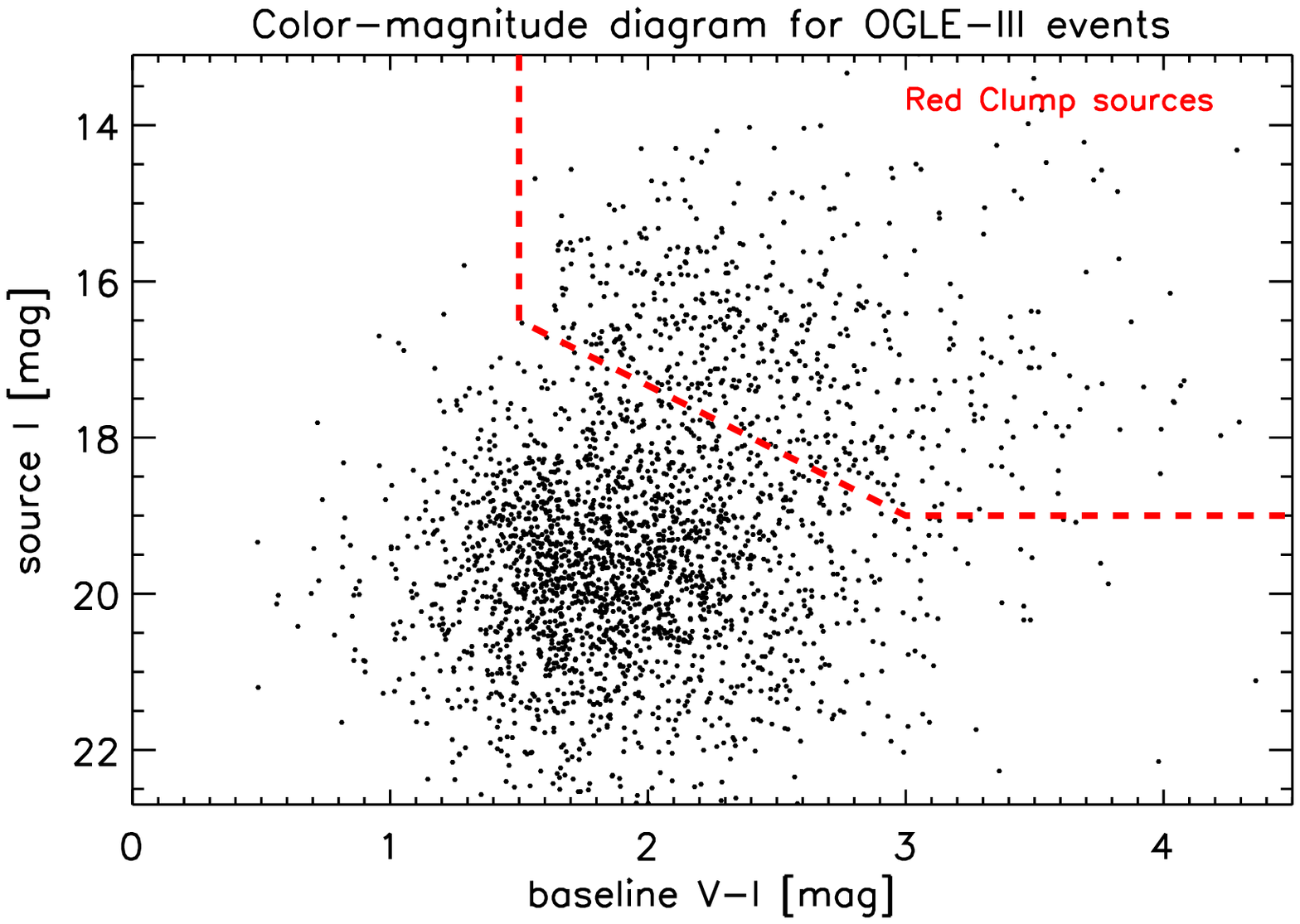}
\caption{Color-Magnitude Diagram of sources in All Sample microlensing events, composed of source $I$-band magnitude computed from the blended microlensing fit and approximate $V-I$ from the baseline. Region with sources belonging to the bulge Red Clump population is marked.}
\label{fig:cmd}
\end{figure}

\begin{figure}
\includegraphics[width=8cm]{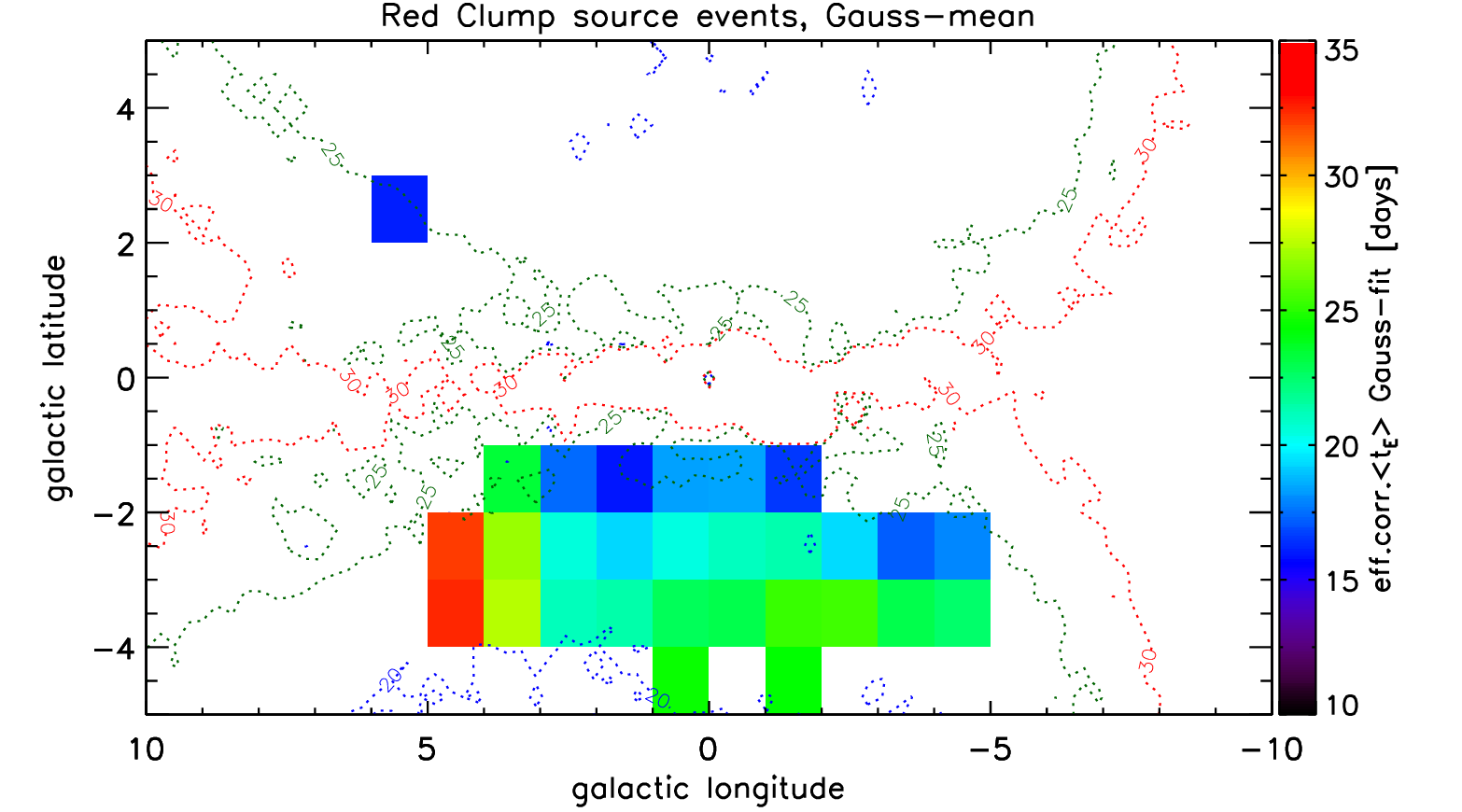}
\caption{
Map of efficiency corrected mean $\tE$ from the log-normal model fit to the $\tE$ distribution in each bin. 
Map was build for 556 events with Red Clump stars from the bulge as most likely sources.
Dashed contours show the expected mean time-scale for events with $\IS<19$ mag \citep{Kerins2009}, shown here for reference only. Bin size is 1$\times$1 degrees.
}
\label{fig:temapRC}
\end{figure}

We can also compare our map of standard mean time-scales for all class A events to the results of MOA-II \citep{Sumi2013} (their Fig. 3, upper-most panel), which were based on 474 events. 
First, we see that our longest mean time-scale is about 35 days, the value which actually appears only at the edges of our observing region and should be ignored as induced by very low number statistics in those areas (those regions completely disappear when using log-normal model mean time-scales, see bottom map in Fig.\ref{fig:temapALL}).
Therefore, our longest value to compare with MOA-II map is actually closer to 25 days, whereas the map in \cite{Sumi2013} shows areas with $\langle \tE \rangle$ as high as 40 days. 
Moreover, on our map we do not see any of the regions of high $\langle \tE \rangle$ values, especially at $l\sim +3$, claimed by MOA-II and previous results (\cite{Alcock1997}, \cite{Popowski2005}), indicating it was likely a statistical fluctuation driven by small number of events available for those studies.
The only part where maps from OGLE-III and MOA-II agree is the decrease of mean $\tE$ at $l\approx +1$, $b\approx -2$, which is even more clearly visible on the log-normal fit mean $\tE$ map for OGLE-III. 
However, as mentioned above, the All-sample maps present a blurred view on mean time-scales due to mixing populations and distances of sources and lenses.

\subsubsection{Bulge-source events sample}
In order to minimize the impact on population mixing, we restricted our events to those with Red Clump (RC) stars from the bulge as the most likely sources.
Figure \ref{fig:cmd} shows the color-magnitude diagram of sources of all events, marking the region of Red Clump sources. 
The $I$-band magnitude was computed using $\I0$ and $\fs$, however, because we did not have enough observations taken in the $V$-band, the color of the source is assumed to be similar to the color of the baseline (\ie source and blends).
This approximation is close to reality because Red Clump giants are typically much brighter than other (bluer) stars in the direction of the Bulge, hence the color of the red giant dominates in the baseline. 
Note that the opposite assumption, that the non-RC sources are bluer in the baseline, however, would not be true for the same reason.

Selecting RC sources allowed us to construct another map of mean time-scales, shown in Fig. \ref{fig:temapRC}.
The values of $\langle \tE \rangle$ are clearly much smaller than in the previous maps, most likely because in this case most of events have sources in the bar and lenses are from the disk. 
Such combinations of lenses and sources typically have higher relative proper motion, hence tend to produce shorter $\tE$ (\eg \citealt{DiStefano2012}.
However, Fig. \ref{fig:temapRC}, has a clearly distinctive two regions with different values of mean time-scale, with the split at about $b\sim-3$ deg. 
The events closer to the Galactic Center tend to have shorter $\tE$ (well below 20 days), whereas those at higher Galactic latitudes have $\tE$ above 20 days. 
The duality, or a gradient of $\langle \tE \rangle$ with galactic latitudes, can be explained that at low $b$ the dominating lensing configuration contains both a source and a lens from the Galactic bar, hence causing shorter events due to larger velocity dispersion.
On the other hand, at larger $b$ the bar density decreases and we relative fraction between bar-bar and bar-disk events changes in favor of bar-disk events, which tend to be longer. 

Our values of $\langle \tE \rangle$ for the Red Clump sample do not match either of two scenarios presented in Fig.5 of \cite{EvansBelokurov2002}, showing mean time-scales for bar sources in Freudenreich's model including and excluding the contribution of bar streaming. 
It might mean that there is very little streaming in the bar, however, our sample of RC events is small and is distributed over a small area of the sky, which might be influencing the conclusions.
More microlensing data is needed for larger galactic longitudes to verify this issue.

\subsubsection{Mass Function}
\begin{figure*}
\center
\includegraphics[width=12cm]{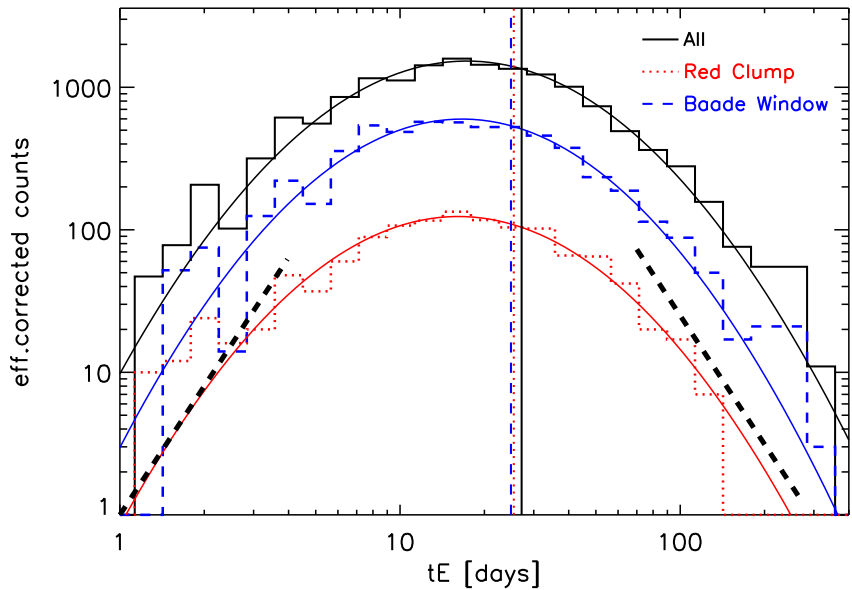}
\caption{Distribution of the logarithm of the Einstein Radius crossing times ($\tE$) for All events (black solid), Red Clump source events (red dotted) and around the Baade's Window (blue dashed).  
We only show events for which the relative error on the time-scale is smaller than 100\%. 
There are 3500,  556 and 1315 events in All, RC and BW samples, respectively. 
The log-normal model fit mean values of $\tE$ are marked with corresponding vertical lines (27.2,  25.5, 24.9 days, respectively).
Thick dashed lines indicate power-law slopes of +3 and -3 for short and long time-scale tails, respectively.
}
\label{fig:te}
\end{figure*}

Finally, in Fig. \ref{fig:te} we show the distributions of efficiency corrected time-scales of events in three subgroups: All, Red Clump and within the 2 degrees around the Baade's Window (BW, $l\sim 1$, $b\sim -2$).
Thick dashed lines show the expected power-law slopes for short and long tail of the distribution of the time-scales, valid for a range of mass functions from \cite{MaoPaczynski1996}.
The slope of the distribution for long end agrees with the prediction, at least for the Red Clump sources. 
For the BW and All events, the long event tail seems to have more shallow slope, which could be caused by either flatter mass function of the disk, or small relative velocity dispersion for disk-disk events, as the BW and All samples contain more disk-disk events, which have typically longer $\tE$. 
This could also just be the modeling bias, mentioned earlier (see sec.\ref{sec:properties}).
For the short time-scales the slope in the data is more flat, which could be a sign of some additional population of either light or fast lenses \citep{Sumi2011}. 
Log-normal model fit average values for $\tE$ were computed as 27.2, 25.5 and 24.9 days for All, RC and BW samples, respectively. 


\subsubsection{Very long and very short events}
We note that our sample contains a significant number of long and short events: there are 9 events with $\tE$ longer than 300d, 205 (7\%) events longer than 100d and 722 (24\%) longer than 50d.
Because our search pipeline was designed to select purely standard events, the events with strong parallax signal were excluded, however, still there remains a couple of events exhibiting some weak systematic deviation from the standard microlensing model likely due to the parallax effect. 
The difference in $\tE$ with and without parallax model included in those cases was within the reported error-bars for $\tE$, however. Therefore, those events can be safely included in our sample of standard events. 
The entire OGLE-III data will be searched separately for long events with strong parallax.

For short events, we see that 561 (19\%) events have $\tE<10$d, 118 (4\%) have $\tE<5$d and there are even 6 events with time-scales between 1 and 2 days. 
Given a relatively sparse sampling of the OGLE-III survey and low detection efficiency for very short events, this might represent a high abundance of low-mass lensing objects like brown dwarfs or unbound or wide-orbit planets, or just a surplus of very fast lenses.
A systematic optimized search for very short OGLE-III and OGLE-IV events will be conducted in near future, along with the study of potential instrumental or numerical systematics causing the time-scales becoming shorter than in reality.

\section{Summary}

We have prepared and presented the largest homogenous catalog of 3718 good quality standard microlensing events toward the Galactic Bulge using 8 years of the OGLE-III survey (2001-2009). 
1409 of the events are new and were not previously reported by the OGLE EWS. 
The full catalog is available for the astronomical community, with all the light curves, probability distributions plots for the microlensing parameters and values of all parameters with their uncertainties in a machine-readable form. 

Applications of such catalog are numerous, from studying the individual events, to large sample analyzes. 
Because of their number and distribution over a wide range of galactic longitudes and latitudes, it was possible to measure the mean $\tE$ per region of the sky robustly enough to provide a direct comparison to the expectations. 
Our data are in good agreement with recent models of the inner parts of the Milky Way, except that the asymmetry in the mean time-scale due to the bar angle seems to be somewhat more pronounced. 
This might be an evidence of a slightly different geometry or shape of the Galactic bar and may call for new models of the very center of our Galaxy.
In near future, the catalog will be the basis for the optical depth and microlensing rate computation for the OGLE-III data, which will further allow for the verification of the models of the Milky Way.

OGLE-IV survey during its first three years of operation has already detected about 5000 events, reported by the EWS. 
An analysis similar to the one presented here on the archival OGLE-IV data will supplement the catalog of events from the OGLE-III and form a colossal sample of standard microlensing events, spread over a wide range of galactic coordinates, allowing for very detailed studies of the inner parts of the Galaxy. 
OGLE-IV is hoped to continue its functioning for many years ahead hence there is room for synergy with the ESA's corner stone astrometric mission Gaia. 
This will create a first time ever opportunity to combine astrometry and photometry and derive masses and distances to individual lenses in microlensing events (\citealt{BelokurovEvans2002}, \citealt{Wyrzykowski2012}), bringing a completely new perspective to studies of the Galactic structure.

\section*{Acknowledgments}
We thank Marzena {\'S}niegowska and Mateusz Zieli{\'n}ski for their contributions during their summer project.
We also acknowledge Drs Nicholas Rattenbury, Eamonn Kerins and Shude Mao for their useful comments and help. 
We thank anonymous referee for their comments and corrections, which improved the paper.

{\L}W and AER acknowledge support from the Polish Ministry of Science under "Iuventus Plus" programme, grant number 9003/IP3/2012/71.
The OGLE project has received funding from the European Research Council under the European CommunityÕs Seventh Framework Programme (FP7/2007-2013)/ERC grant agreement no. 246678 to AU.

%

\end{document}

%% file: table1.tex
0001 & 17:51:44.00 & -30:17:20.7 & BLG100.1 & 27898 & 2007-BLG-258 & - \\
0002 & 17:52:21.54 & -30:16:19.0 & BLG100.1 & 47983 & - & - \\
0003 & 17:51:01.18 & -30:15:28.9 & BLG100.1 & 55988 & 2007-BLG-459 & - \\
0004 & 17:51:11.06 & -30:14:20.6 & BLG100.1 & 57414 & - & - \\
0005 & 17:51:43.85 & -30:14:43.3 & BLG100.1 & 89325 & - & - \\
0006 & 17:52:13.46 & -30:13:22.9 & BLG100.1 & 101269 & 2005-BLG-404 & - \\
0007 & 17:52:08.57 & -30:14:56.3 & BLG100.1 & 102672 & - & - \\
0008 & 17:52:19.37 & -30:13:13.9 & BLG100.1 & 106173 & 2005-BLG-294 & - \\
0009 & 17:51:16.18 & -30:12:31.9 & BLG100.1 & 119596 & 2002-BLG-001 & - \\
0010 & 17:51:16.71 & -30:11:10.7 & BLG100.1 & 124027 & - & - \\
0011 & 17:52:13.40 & -30:12:32.8 & BLG100.1 & 160378 & 2004-BLG-435 & - \\
0012 & 17:51:17.94 & -30:09:12.7 & BLG100.1 & 180707 & 2007-BLG-602 & - \\
0013 & 17:51:29.14 & -30:09:13.4 & BLG100.1 & 186215 & 2007-BLG-532 & - \\
0014 & 17:51:34.36 & -30:10:07.3 & BLG100.1 & 192835 & 2004-BLG-484 & - \\
0015 & 17:51:47.89 & -30:09:17.7 & BLG100.1 & 198644 & 2005-BLG-251 & - \\
0016 & 17:52:03.54 & -30:10:42.4 & BLG100.1 & 206957 & - & - \\
0017 & 17:52:08.26 & -30:09:44.1 & BLG100.1 & 214894 & - & - \\
0018 & 17:51:12.37 & -30:05:04.1 & BLG100.2 & 70077 & 2004-BLG-153 & - \\
0019 & 17:51:43.83 & -30:04:59.5 & BLG100.2 & 91698 & 2004-BLG-229 & - \\
0020 & 17:52:03.51 & -30:05:32.6 & BLG100.2 & 97813 & - & - \\
0021 & 17:51:01.97 & -30:02:00.2 & BLG100.2 & 119749 & 2004-BLG-115 & - \\
0022 & 17:52:09.43 & -30:03:01.3 & BLG100.2 & 164216 & 2008-BLG-303 & - \\
0023 & 17:52:13.07 & -30:03:30.3 & BLG100.2 & 164721 & - & - \\
0024 & 17:51:28.62 & -30:01:00.2 & BLG100.2 & 188384 & - & - \\
0025 & 17:51:48.79 & -30:00:15.9 & BLG100.2 & 201309 & 2004-BLG-192 & - \\
0026 & 17:51:05.19 & -29:58:58.9 & BLG100.3 & 1858 & 2006-BLG-097 & - \\
0027 & 17:51:35.90 & -29:59:08.8 & BLG100.3 & 21115 & - & - \\
0028 & 17:51:20.69 & -29:55:52.1 & BLG100.3 & 63514 & 2004-BLG-167 & - \\
0029 & 17:51:34.32 & -29:55:19.7 & BLG100.3 & 74159 & 2002-BLG-193 & - \\
0030 & 17:52:06.70 & -29:55:16.2 & BLG100.3 & 98639 & 2004-BLG-028 & - \\
0031 & 17:51:29.95 & -29:54:19.7 & BLG100.3 & 123110 & 2007-BLG-250 & - \\
0032 & 17:51:25.78 & -29:53:50.6 & BLG100.3 & 123957 & 2004-BLG-034 & - \\
0033 & 17:52:01.31 & -29:53:56.2 & BLG100.3 & 144687 & 2004-BLG-281 & - \\
0034 & 17:52:00.25 & -29:54:39.7 & BLG100.3 & 147281 & 2007-BLG-166 & - \\
0035 & 17:51:05.26 & -29:52:11.6 & BLG100.3 & 167671 & 2006-BLG-018 & - \\
0036 & 17:52:14.55 & -29:50:51.9 & BLG100.3 & 210689 & 2007-BLG-219 & - \\
0037 & 17:51:32.62 & -29:48:38.2 & BLG100.4 & 17039 & 2004-BLG-135 & - \\
0038 & 17:51:38.28 & -29:47:16.3 & BLG100.4 & 73670 & 2005-BLG-172 & - \\
0039 & 17:52:16.62 & -29:44:25.7 & BLG100.4 & 159146 & 2006-BLG-098 & - \\
0040 & 17:51:42.83 & -29:42:44.0 & BLG100.4 & 187869 & 2008-BLG-454 & - \\
0041 & 17:52:15.81 & -29:42:16.4 & BLG100.4 & 206107 & 2006-BLG-019 & - \\
0042 & 17:50:48.09 & -29:49:30.3 & BLG100.5 & 29000 & - & - \\
0043 & 17:49:38.48 & -29:55:46.5 & BLG100.6 & 37372 & - & - \\
0044 & 17:49:47.24 & -29:56:42.2 & BLG100.6 & 38608 & 2004-BLG-131 & - \\
0045 & 17:50:33.56 & -29:56:15.4 & BLG100.6 & 57621 & 2008-BLG-144 & - \\
0046 & 17:49:49.54 & -29:54:05.4 & BLG100.6 & 79092 & 2009-BLG-093 & - \\
0047 & 17:50:13.79 & -29:52:42.1 & BLG100.6 & 87913 & 2004-BLG-301 & - \\
0048 & 17:50:09.90 & -29:52:59.0 & BLG100.6 & 91567 & 2007-BLG-440 & - \\
0049 & 17:50:25.92 & -29:52:48.4 & BLG100.6 & 93188 & 2008-BLG-265 & - \\

%% file: table2.tex
0001 & BLG100.1.27898 & 4247.210$_{-0.003}^{+0.003}$ & 2.58$_{-0.10}^{+0.12}$ & 0.079670$_{-0.011984}^{+0.010063}$ & 0.946$_{-0.067}^{+0.063}$ & 17.817$_{-0.001}^{+0.001}$ & 5321.16 & 2384 & 20.932$\pm$0.101 \\
0002 & BLG100.1.47983 & 3276.397$_{-0.227}^{+0.201}$ & 53.84$_{-10.89}^{+11.90}$ & 0.071218$_{-0.016150}^{+0.025514}$ & 0.028$_{-0.006}^{+0.010}$ & 16.766$_{-0.000}^{+0.000}$ & 3586.67 & 2173 & 18.257$\pm$0.018 \\
0003 & BLG100.1.55988 & 4327.090$_{-0.602}^{+0.610}$ & 14.80$_{-1.71}^{+4.15}$ & 0.664801$_{-0.126552}^{+0.225230}$ & 0.761$_{-0.356}^{+0.269}$ & 18.818$_{-0.002}^{+0.002}$ & 1934.39 & 1515 & 99.999$\pm$9.999 \\
0004 & BLG100.1.57414 & 4230.126$_{-0.030}^{+0.027}$ & 6.29$_{-1.26}^{+1.78}$ & 0.148585$_{-0.042069}^{+0.051833}$ & 0.465$_{-0.141}^{+0.189}$ & 18.801$_{-0.002}^{+0.002}$ & 3311.50 & 2379 & 21.242$\pm$0.260 \\
0005 & BLG100.1.89325 & 3577.832$_{-0.013}^{+0.016}$ & 2.24$_{-0.39}^{+0.42}$ & 0.049491$_{-0.013952}^{+0.010520}$ & 0.726$_{-0.144}^{+0.222}$ & 19.992$_{-0.005}^{+0.005}$ & 2689.08 & 2382 & 99.999$\pm$9.999 \\
0006 & BLG100.1.101269 & 3576.846$_{-0.002}^{+0.002}$ & 10.29$_{-0.42}^{+0.44}$ & 0.017707$_{-0.000776}^{+0.000837}$ & 0.557$_{-0.025}^{+0.025}$ & 18.879$_{-0.002}^{+0.002}$ & 4255.68 & 2388 & 99.999$\pm$9.999 \\
0007 & BLG100.1.102672 & 3466.213$_{-0.575}^{+0.613}$ & 12.22$_{-1.98}^{+6.06}$ & 0.400672$_{-0.196534}^{+0.740753}$ & 0.713$_{-0.403}^{+0.347}$ & 19.378$_{-0.003}^{+0.003}$ & 3653.14 & 2387 & 99.999$\pm$9.999 \\
0008 & BLG100.1.106173 & 3537.583$_{-0.081}^{+0.090}$ & 8.02$_{-0.34}^{+0.48}$ & 0.517930$_{-0.045981}^{+0.037340}$ & 0.869$_{-0.114}^{+0.101}$ & 17.192$_{-0.000}^{+0.000}$ & 2838.19 & 2374 & 20.001$\pm$0.065 \\
0009 & BLG100.1.119596 & 2380.609$_{-0.130}^{+0.152}$ & 66.80$_{-0.88}^{+1.15}$ & 0.365873$_{-0.008614}^{+0.011041}$ & 0.410$_{-0.016}^{+0.013}$ & 15.495$_{-0.000}^{+0.000}$ & 5805.34 & 2388 & 19.131$\pm$0.031 \\
0010 & BLG100.1.124027 & 3120.437$_{-0.296}^{+0.277}$ & 12.80$_{-1.29}^{+2.44}$ & 0.320332$_{-0.085190}^{+0.057526}$ & 1.038$_{-0.289}^{+0.218}$ & 19.805$_{-0.004}^{+0.004}$ & 2921.23 & 2382 & 99.999$\pm$9.999 \\
0011 & BLG100.1.160378 & 3213.992$_{-0.071}^{+0.069}$ & 24.78$_{-3.32}^{+3.78}$ & 0.094060$_{-0.027038}^{+0.021788}$ & 0.469$_{-0.087}^{+0.113}$ & 18.978$_{-0.002}^{+0.002}$ & 3041.73 & 2388 & 99.999$\pm$9.999 \\
0012 & BLG100.1.180707 & 4414.838$_{-1.232}^{+1.879}$ & 43.75$_{-3.67}^{+6.51}$ & 0.528626$_{-0.077126}^{+0.124027}$ & 0.876$_{-0.260}^{+0.188}$ & 18.870$_{-0.002}^{+0.002}$ & 2197.70 & 1982 & 99.999$\pm$9.999 \\
0013 & BLG100.1.186215 & 4350.822$_{-0.078}^{+0.075}$ & 9.14$_{-1.92}^{+3.17}$ & 0.166970$_{-0.065877}^{+0.054707}$ & 0.590$_{-0.211}^{+0.282}$ & 19.062$_{-0.002}^{+0.003}$ & 2958.23 & 2085 & 99.999$\pm$9.999 \\
0014 & BLG100.1.192835 & 3230.394$_{-0.014}^{+0.014}$ & 9.26$_{-0.68}^{+0.78}$ & 0.110588$_{-0.011216}^{+0.010884}$ & 0.792$_{-0.081}^{+0.087}$ & 18.520$_{-0.001}^{+0.001}$ & 2818.92 & 2386 & 99.999$\pm$9.999 \\
0015 & BLG100.1.198644 & 3516.324$_{-0.053}^{+0.065}$ & 4.48$_{-0.25}^{+0.37}$ & 0.000101$_{-0.071132}^{+0.070935}$ & 0.888$_{-0.125}^{+0.114}$ & 17.909$_{-0.001}^{+0.001}$ & 3164.87 & 2300 & 20.685$\pm$0.059 \\
0016 & BLG100.1.206957 & 4318.948$_{-0.096}^{+0.113}$ & 5.92$_{-1.09}^{+1.59}$ & 0.324206$_{-0.123125}^{+0.100714}$ & 0.576$_{-0.206}^{+0.314}$ & 18.372$_{-0.001}^{+0.001}$ & 5094.66 & 2387 & 99.999$\pm$9.999 \\
0017 & BLG100.1.214894 & 4549.661$_{-0.724}^{+0.664}$ & 16.36$_{-3.10}^{+7.35}$ & 0.577917$_{-0.258365}^{+0.219489}$ & 0.550$_{-0.309}^{+0.401}$ & 19.125$_{-0.002}^{+0.002}$ & 2921.61 & 2323 & 99.999$\pm$9.999 \\
0018 & BLG100.2.70077 & 3110.076$_{-0.178}^{+0.158}$ & 17.28$_{-2.88}^{+3.58}$ & 0.202331$_{-0.046873}^{+0.061412}$ & 0.842$_{-0.216}^{+0.322}$ & 19.854$_{-0.004}^{+0.004}$ & 2686.07 & 2383 & 99.999$\pm$9.999 \\
0019 & BLG100.2.91698 & 3137.961$_{-0.078}^{+0.073}$ & 11.50$_{-2.52}^{+3.05}$ & 0.105307$_{-0.029340}^{+0.041323}$ & 0.212$_{-0.057}^{+0.092}$ & 18.670$_{-0.001}^{+0.001}$ & 3065.78 & 2388 & 99.999$\pm$9.999 \\
0020 & BLG100.2.97813 & 4922.779$_{-1.230}^{+1.178}$ & 41.16$_{-6.77}^{+12.81}$ & 0.721773$_{-0.260755}^{+0.199362}$ & 0.120$_{-0.061}^{+0.069}$ & 17.004$_{-0.000}^{+0.000}$ & 3815.13 & 2388 & 18.787$\pm$0.016 \\
0021 & BLG100.2.119749 & 3094.519$_{-0.097}^{+0.094}$ & 9.53$_{-1.22}^{+1.40}$ & 0.458382$_{-0.129482}^{+0.098796}$ & 0.513$_{-0.139}^{+0.224}$ & 17.078$_{-0.000}^{+0.000}$ & 19945.97 & 2085 & 21.588$\pm$0.223 \\
0022 & BLG100.2.164216 & 4635.154$_{-0.017}^{+0.015}$ & 37.97$_{-1.65}^{+1.17}$ & 0.015255$_{-0.001601}^{+0.001525}$ & 0.738$_{-0.026}^{+0.040}$ & 19.232$_{-0.002}^{+0.002}$ & 3269.30 & 2388 & 99.999$\pm$9.999 \\
0023 & BLG100.2.164721 & 3795.421$_{-0.082}^{+0.077}$ & 6.03$_{-0.46}^{+0.88}$ & 0.288698$_{-0.036700}^{+0.059969}$ & 0.902$_{-0.215}^{+0.146}$ & 18.655$_{-0.002}^{+0.002}$ & 6201.02 & 2388 & 99.999$\pm$9.999 \\
0024 & BLG100.2.188384 & 3912.878$_{-0.024}^{+0.026}$ & 5.84$_{-0.26}^{+0.35}$ & 0.265964$_{-0.018859}^{+0.021966}$ & 0.963$_{-0.098}^{+0.086}$ & 18.646$_{-0.001}^{+0.001}$ & 2979.01 & 2387 & 99.999$\pm$9.999 \\
0025 & BLG100.2.201309 & 3135.172$_{-0.121}^{+0.118}$ & 12.81$_{-1.65}^{+1.91}$ & 0.282024$_{-0.052400}^{+0.067166}$ & 0.594$_{-0.129}^{+0.186}$ & 18.349$_{-0.001}^{+0.001}$ & 3720.61 & 2344 & 99.999$\pm$9.999 \\
0026 & BLG100.3.1858 & 3819.778$_{-0.030}^{+0.028}$ & 21.90$_{-2.27}^{+3.77}$ & 0.051051$_{-0.007149}^{+0.008610}$ & 0.700$_{-0.113}^{+0.100}$ & 19.953$_{-0.014}^{+0.015}$ & 978.58 & 653 & 99.999$\pm$9.999 \\
0027 & BLG100.3.21115 & 3194.346$_{-0.456}^{+0.533}$ & 40.42$_{-11.14}^{+14.43}$ & 0.118756$_{-0.039454}^{+0.066169}$ & 0.116$_{-0.039}^{+0.075}$ & 19.141$_{-0.002}^{+0.002}$ & 2402.90 & 2385 & 99.999$\pm$9.999 \\
0028 & BLG100.3.63514 & 3115.796$_{-0.118}^{+0.124}$ & 5.65$_{-1.09}^{+2.26}$ & 0.381420$_{-0.144852}^{+0.151985}$ & 0.587$_{-0.278}^{+0.346}$ & 18.775$_{-0.001}^{+0.001}$ & 2853.57 & 2388 & 21.569$\pm$0.184 \\
0029 & BLG100.3.74159 & 2443.201$_{-0.072}^{+0.061}$ & 38.49$_{-5.80}^{+6.59}$ & 0.002225$_{-0.011397}^{+0.016117}$ & 0.011$_{-0.002}^{+0.003}$ & 16.609$_{-0.000}^{+0.000}$ & 4637.86 & 2388 & 20.088$\pm$0.076 \\
0030 & BLG100.3.98639 & 3065.941$_{-0.544}^{+0.553}$ & 22.59$_{-2.18}^{+4.66}$ & 0.448075$_{-0.123493}^{+0.072495}$ & 0.862$_{-0.299}^{+0.211}$ & 19.014$_{-0.002}^{+0.002}$ & 4440.65 & 2388 & 99.999$\pm$9.999 \\
0031 & BLG100.3.123110 & 4243.517$_{-0.096}^{+0.090}$ & 6.63$_{-1.49}^{+1.52}$ & 0.408378$_{-0.183294}^{+0.101790}$ & 0.160$_{-0.052}^{+0.121}$ & 16.800$_{-0.000}^{+0.000}$ & 3391.71 & 2388 & 20.080$\pm$0.067 \\
0032 & BLG100.3.123957 & 3088.500$_{-0.327}^{+0.298}$ & 25.92$_{-2.55}^{+3.44}$ & 0.493918$_{-0.088542}^{+0.086958}$ & 0.762$_{-0.191}^{+0.220}$ & 18.271$_{-0.001}^{+0.001}$ & 2504.46 & 2388 & 99.999$\pm$9.999 \\
0033 & BLG100.3.144687 & 3155.836$_{-0.325}^{+0.302}$ & 281.12$_{-76.07}^{+72.71}$ & 0.018223$_{-0.007665}^{+0.004056}$ & 0.002$_{-0.000}^{+0.001}$ & 15.804$_{-0.000}^{+0.000}$ & 4693.40 & 2379 & 19.976$\pm$0.077 \\
0034 & BLG100.3.147281 & 4209.446$_{-0.062}^{+0.065}$ & 9.94$_{-1.37}^{+1.87}$ & 0.080589$_{-0.021768}^{+0.025282}$ & 0.585$_{-0.125}^{+0.149}$ & 19.182$_{-0.002}^{+0.002}$ & 3195.60 & 2388 & 99.999$\pm$9.999 \\
0035 & BLG100.3.167671 & 3802.265$_{-0.047}^{+0.044}$ & 19.23$_{-0.91}^{+0.85}$ & 0.268731$_{-0.020916}^{+0.017090}$ & 0.802$_{-0.060}^{+0.079}$ & 17.608$_{-0.001}^{+0.001}$ & 3261.91 & 2379 & 21.338$\pm$0.144 \\

%% file: table3.tex
90001 & 17:52:12.90 & -30:13:38.5 & BLG100.1 & 101182 & 2005-BLG-359\\
90002 & 17:51:19.10 & -30:02:47.0 & BLG100.2 & 132033 & -\\
90003 & 17:50:32.77 & -29:52:13.7 & BLG100.6 & 137109 & 2007-BLG-194\\
90004 & 17:50:06.77 & -30:08:46.5 & BLG100.7 & 9419 & -\\
90005 & 17:53:52.53 & -29:59:55.0 & BLG101.1 & 238410 & 2007-BLG-090\\
90006 & 17:52:43.65 & -29:39:20.2 & BLG101.5 & 20164 & -\\
90007 & 17:52:45.56 & -29:49:46.9 & BLG101.6 & 22645 & -\\
90008 & 17:53:29.42 & -29:56:57.0 & BLG101.7 & 59322 & -\\
90009 & 17:53:24.17 & -29:56:32.7 & BLG101.7 & 124729 & -\\
90010 & 17:55:55.53 & -29:21:14.5 & BLG102.5 & 46864 & 2008-BLG-552\\
90011 & 17:55:13.27 & -29:16:53.1 & BLG102.5 & 135824 & -\\
90012 & 17:55:04.70 & -29:25:14.0 & BLG102.6 & 139751 & -\\
90013 & 17:55:30.55 & -29:24:39.4 & BLG102.6 & 161973 & 2007-BLG-295\\
90014 & 17:55:50.73 & -29:45:40.6 & BLG102.8 & 108332 & -\\
90015 & 17:56:47.26 & -30:01:28.4 & BLG103.3 & 151443 & 2006-BLG-053\\
90016 & 17:56:24.60 & -29:50:19.7 & BLG103.4 & 124317 & -\\
90017 & 17:55:14.38 & -29:53:14.0 & BLG103.5 & 79851 & -\\
90018 & 17:56:05.22 & -30:11:04.9 & BLG103.7 & 113081 & -\\
90019 & 17:55:11.11 & -30:22:04.3 & BLG103.8 & 54768 & 2004-BLG-486, 2004-BLG-490\\
90020 & 17:59:23.08 & -29:45:01.5 & BLG104.1 & 17566 & -\\
90021 & 17:59:47.01 & -29:18:13.3 & BLG104.4 & 35755 & -\\
90022 & 17:59:16.67 & -29:15:14.1 & BLG104.4 & 76137 & -\\
90023 & 17:59:43.35 & -29:14:37.8 & BLG104.4 & 105201 & -\\
90024 & 17:57:53.42 & -29:12:03.9 & BLG104.5 & 194584 & -\\
90025 & 17:58:40.44 & -29:11:52.7 & BLG104.5 & 225289 & 2006-BLG-202\\
90026 & 18:02:28.53 & -29:43:26.2 & BLG105.1 & 105229 & -\\
90027 & 18:01:13.71 & -29:27:30.1 & BLG105.6 & 40335 & 2002-BLG-357\\
90028 & 18:01:38.68 & -29:20:28.5 & BLG105.6 & 240417 & -\\
90029 & 17:46:33.26 & -36:42:32.2 & BLG108.2 & 3390 & 2004-BLG-517\\
90030 & 17:46:57.18 & -34:27:57.6 & BLG130.1 & 56220 & -\\
90031 & 17:45:09.46 & -33:56:48.3 & BLG130.5 & 98747 & 2007-BLG-275\\
90032 & 17:46:19.48 & -34:13:45.2 & BLG130.7 & 200260 & 2006-BLG-262\\
90033 & 17:49:19.54 & -34:23:00.9 & BLG131.1 & 141016 & -\\
90034 & 17:48:15.25 & -34:06:53.7 & BLG131.6 & 104924 & 2004-BLG-099\\
90035 & 17:52:40.90 & -34:26:59.4 & BLG132.1 & 72963 & -\\
90036 & 17:53:09.51 & -34:08:26.3 & BLG132.3 & 89578 & -\\
90037 & 17:56:59.79 & -33:55:15.6 & BLG134.5 & 193547 & 2005-BLG-061\\
90038 & 17:45:31.29 & -33:46:15.7 & BLG138.1 & 163835 & -\\
90039 & 17:46:14.66 & -33:47:27.9 & BLG138.1 & 185675 & -\\
90040 & 17:44:55.31 & -33:25:03.2 & BLG138.5 & 70885 & 2008-BLG-238\\
90041 & 17:45:03.92 & -33:52:29.2 & BLG138.8 & 81987 & 2004-BLG-499\\
90042 & 17:47:05.93 & -33:40:46.9 & BLG139.7 & 75690 & 2006-BLG-054\\
90043 & 17:50:42.23 & -33:30:57.8 & BLG140.6 & 138482 & 2007-BLG-476\\
90044 & 17:57:00.80 & -33:53:58.7 & BLG142.1 & 24051 & 2007-BLG-311\\
90045 & 17:51:05.78 & -32:45:47.4 & BLG147.4 & 157489 & -\\
90046 & 17:52:07.67 & -33:11:44.1 & BLG148.8 & 193131 & 2006-BLG-363\\
90047 & 17:53:51.79 & -33:14:12.4 & BLG149.8 & 72373 & 2004-BLG-567\\
90048 & 17:57:23.66 & -33:06:11.0 & BLG150.7 & 75044 & 2004-BLG-016\\
90049 & 17:54:51.71 & -32:21:46.7 & BLG156.6 & 116230 & 2008-BLG-496\\
90050 & 17:54:04.49 & -32:40:38.2 & BLG156.8 & 68730 & -\\

%% file: table4.tex
5.5 & -3.5 & 34.4 & 32.8 & 31.9 & 9 \\
5.5 & 2.5 & 46.8 & 37.5 & 30.9 & 10 \\
4.5 & -3.5 & 47.1 & 42.5 & 23.4 & 17 \\
4.5 & -2.5 & 42.8 & 32.2 & 29.4 & 18 \\
3.5 & -4.5 & 29.1 & 24.1 & 19.7 & 18 \\
3.5 & -3.5 & 33.5 & 24.0 & 19.7 & 39 \\
3.5 & -2.5 & 34.9 & 27.2 & 18.6 & 33 \\
3.5 & -1.5 & 30.8 & 15.6 & 24.5 & 11 \\
2.5 & -4.5 & 28.5 & 22.4 & 23.1 & 19 \\
2.5 & -3.5 & 26.7 & 20.3 & 17.5 & 40 \\
2.5 & -2.5 & 23.7 & 16.5 & 12.4 & 75 \\
2.5 & -1.5 & 29.4 & 24.4 & 15.5 & 18 \\
1.5 & -4.5 & 29.5 & 26.4 & 22.1 & 23 \\
1.5 & -3.5 & 30.1 & 24.6 & 21.2 & 38 \\
1.5 & -2.5 & 26.9 & 20.8 & 19.9 & 98 \\
1.5 & -1.5 & 15.6 & 11.4 & 12.7 & 20 \\
0.5 & -4.5 & 27.8 & 17.4 & 28.0 & 17 \\
0.5 & -3.5 & 28.0 & 24.3 & 23.8 & 53 \\
0.5 & -2.5 & 26.0 & 19.3 & 16.6 & 92 \\
0.5 & -1.5 & 31.8 & 25.5 & 19.5 & 97 \\
0.5 & 2.5 & 27.9 & 22.4 & 15.1 & 6 \\
-0.5 & -4.5 & 22.4 & 20.8 & 11.4 & 12 \\
-0.5 & -3.5 & 21.6 & 17.2 & 16.0 & 35 \\
-0.5 & -2.5 & 26.4 & 18.3 & 22.4 & 56 \\
-0.5 & -1.5 & 24.4 & 19.3 & 17.4 & 47 \\
-1.5 & -4.5 & 28.5 & 20.8 & 17.5 & 16 \\
-1.5 & -3.5 & 33.5 & 23.3 & 27.9 & 38 \\
-1.5 & -2.5 & 34.3 & 27.4 & 25.0 & 37 \\
-1.5 & -1.5 & 26.0 & 22.1 & 18.0 & 15 \\
-2.5 & -4.5 & 46.8 & 35.0 & 16.3 & 10 \\
-2.5 & -3.5 & 38.7 & 29.1 & 28.9 & 31 \\
-2.5 & -2.5 & 32.0 & 24.4 & 16.1 & 17 \\
-3.5 & -3.5 & 27.0 & 22.2 & 23.5 & 16 \\
-3.5 & -2.5 & 28.5 & 18.6 & 36.5 & 12 \\
-4.5 & -3.5 & 39.2 & 33.8 & 23.9 & 28 \\
-4.5 & -2.5 & 35.1 & 28.6 & 19.7 & 32 \\